\documentclass[10pt,twocolumn,english]{revtex4-2}
\usepackage{mathptmx}

\usepackage[T1]{fontenc}
\usepackage[latin9]{inputenc}
\usepackage[letterpaper]{geometry}
\usepackage{xcolor}
\usepackage{textcomp}
\usepackage{amsmath}
\usepackage{amssymb}
\usepackage{cancel}
\usepackage{graphicx}
\usepackage{esint}
\usepackage{bm}
\PassOptionsToPackage{normalem}{ulem}
\usepackage{ulem}
\usepackage[latin9]{inputenc}
\makeatletter

\newcommand{\lyxdot}{.}

\makeatother

\makeatletter

\providecolor{lyxadded}{rgb}{0,0,1}
\providecolor{lyxdeleted}{rgb}{1,0,0}

\DeclareRobustCommand{\lyxsout}[1]{\ifx\\#1\else\sout{#1}\fi}

\usepackage{babel}

\makeatother

\usepackage{babel}
\begin{document}
\title{Thermally Activated Transitions Between Micromagnetic States}
\begin{abstract}
We review work by the authors on thermal activation in nanoscopic magnetic systems. These systems present 
unique difficulties in analyzing noise-induced escape over a barrier, including the presence of nonlocal interactions, nongradient
terms in the energy functional, and dynamical textures as initial or saddle states. We begin with a discussion of magnetic reversal
between single-domain configurations of the magnetization. Here the transition (saddle) state can be either a single-domain or
a spatially varying (instanton-like) configuration, and depending on the system parameters can exhibit either Arrhenius or non-Arrhenius 
reversal rates. We then turn to a discussion of transitions between magnetic textures, which can be either static and topologically protected or dynamic and not topologically protected. An example of the latter case is the droplet soliton, a rotating nontopologically-protected configuration, which we find can occur either as a metastable
or transition state in a nanoscopic magnetic system. After discussing various issues in calculating transition rates, we present results for the activation barriers for creation and annihilation of these magnetic textures.  We conclude with a discussion of activated transitions between topologically protected skyrmion textures and other configurations, on which work is ongoing.
\end{abstract}

\author{Gabriel D.~Chaves-O'Flynn}
\affiliation{Institute of Molecular Physics, Polish Academy of Sciences}
\author{D.L.~Stein}
\affiliation{Department of Physics and Courant Institute of Mathematical Sciences,
New York University, New York, NY 10012 USA}
\affiliation{NYU-ECNU Institutes of Physics and Mathematical Sciences at NYU Shanghai,
3663 Zhongshan Road North, Shanghai, 200062, China}
\affiliation{Santa Fe Institute, 1399 Hyde Park Rd., Santa Fe, NM USA 87501}
\maketitle

\section{Introduction and Dedication}
\label{sec:intro}

Charlie Doering was one of a rare breed, accomplished both as a mathematician and physicist and enormously productive in both fields. 
He was also a wonderful human being --- cheerful, open, energetic, generous, and a true friend. 
His passing was a tremendous loss for his family, friends, and the larger mathematics and physics communities. His scientific legacy, however, will
live on far into the future.

Charlie was perhaps best known for his numerous deep and incisive contributions to the science of turbulence. But his interests spanned a 
broad range of topics in applied mathematics, particularly in problems relating to stochastic processes. It was in this area that one of us (DLS) interacted 
and collaborated with Charlie, leading to a couple of papers investigating what was then the novel problem of stochastic escape over a fluctuating barrier~\cite{SPHD89,SPDHM90}.
These served as a basis for Charlie's best-known work in this area: an investigation into the connection between fluctuating barriers 
and stochastic resonance~\cite{DG92,Maddox92}, which has fruitful applications to problems of molecular transport in biological cells, among others.

Another area in which stochastic methods are important is magnetic systems, particularly information storage and transport in nanomagnets. Because of the
small size of these systems, both quantum and thermal fluctuations can be significant. These fluctuations present both a problem and an opportunity: they can erase stored information due to reversal of magnetic moments,
but they can also be harnessed to lower electrical resistance and assist transport. Although Charlie did not work in this particular area, we are confident he would have enjoyed seeing these methods put to work in an area of both physical significance and practical utility, and it is with this thought that we dedicate this paper to Charlie's memory.

\section{Dynamics and thermal fluctuations in magnetic systems}
\label{sec:flucs}

The modern approach applying Kramers' theory~\cite{Kramers40} to thermal nucleation in magnetic systems dates back to N\'eel~\cite{Neel49} and Brown~\cite{Brown63}, who investigated magnetization reversal in fine ferromagnetic particles comprising a single magnetic domain with uniform magnetization~${\bf m}(t)$. In the absence of random fluctuations, the magnetization obeys the dynamical Landau-Lifschitz-Gilbert~(LLG) equation~\cite{LL35,Gilbert55}
\begin{equation}
\label{eq:LLG}
\frac{\partial{\bf m}}{\partial t}=-\gamma({\bf m}\times {\bf h}_{\rm eff})+\Big(\frac{\alpha}{m_0}\Big)\Big({\bf m}\times\frac{\partial{\bf m}}{\partial t}\Big)
\end{equation}
where $m_0$ is the (fixed) magnetization magnitude, $\alpha>0$ the (dimensionless) phenomenological damping constant, and $\gamma>0$ the gyromagnetic ratio. Typical values for ferromagnetic nanoparticles with radii of order tens of nanometers are $\gamma\approx 10^{11}$T$^{-1}$s$^{-1}$, $\alpha\approx 0.01$,  and $m_0\approx 400-800$ kA$\cdot$m$^{-1}$.
It will be convenient later to define the operator 
\begin{equation}
\label{eq:lop}
{\mathbb L}=-\frac{\gamma m_0}{1+\alpha^2}[{\bf m}\times\ +\ \alpha{\bf m}\times{\bf m}\times]
\end{equation}
so the iterated LLG equation becomes $\dot{\bf m}={\mathbb L}{\bf h}_{\rm eff}$.

As indicated in Eq.~(\ref{eq:LLG}), the dynamics are governed by an effective field~${\bf h}_{\rm eff}=-\delta E/\delta{\bf m}$, the variational derivative of the system energy $E({\bf m})$.
Because we are interested in systems where the magnetization varies in space, we defer a discussion of $E({\bf m})$ to the next section. The more general aspect of the N\'eel-Brown theory, which is applicable to a wide variety of situations, is the incorporation of thermal noise into the magnetization dynamics. We are interested in systems and laboratory conditions where thermal noise strongly dominates quantum effects, which for most systems is for temperatures above a few degrees Kelvin. A simple analysis~\cite{Brown63} demonstrates that in this regime thermal fluctuations can be approximated as white noise added to the effective field~${\bf h}_{\rm eff}$. One can then use the Kramers theory to calculate magnetization reversal times, which follow an Arrhenius law $\tau_{\rm reversal}=\Gamma_0^{-1}\exp[\Delta W/k_BT]$, where the prefactor $\Gamma_0^{-1}$ and the activation energy $\Delta W$ are both independent of temperature. However, as we will see below, magnetic systems present several challenges absent in most other systems.

\section{Magnetization reversal in thin films}
\label{sec:2D}

The classical N\'eel-Brown theory of thermally induced reversal assumed a spatially uniform magnetization and uniaxial anisotropy, and has been experimentally confirmed for simple single-domain systems such as 15-30~nm diameter particles of Ni, Co, and Dy~\cite{Wensdorfer97}. Although the N\'eel-Brown theory works well for spherical nanoparticles, it is not generally applicable to thin films, cylindrical or otherwise elongated magnetic particles, and other geometries, where thermally induced reversal times are orders of magnitude smaller than those predicted by N\'eel-Brown. 

The failure of N\'eel-Brown theory for these systems lies in the assumption of a uniform magnetization. Under this assumption, magnetic reversal requires a rigid rotation of all the spins and so the activation barrier scales linearly with the volume.  For many nonspherical systems, however, the activation energy scales much slower than the volume, indicating for these systems that the uniform magnetization assumption is not valid. As a first step, then, one needs to include spatial variation of the magnetization in the energy function.  In this section we will consider the following Hamiltonian~\cite{MSK05,MSK06}:
\begin{eqnarray}
\label{eq:H}
E[{\bf m}({\bf x})]&=&\lambda^2\int_\Omega d^3x\ \vert\nabla{\bf m}\vert^2+\frac{1}{2}\int_{{\mathbb R}^3}d^3x\ \vert\nabla U\vert^2\nonumber\\
&-&\int_\Omega d^3x\ {\bf h}\cdot{\bf m}
\end{eqnarray}
where the magnetic permeability of the vacuum $\mu_0$ has been set to one, 
$\Omega$ is the region occupied by the ferromagnet, $\lambda$ is the exchange length, $\vert\nabla{\bf m}\vert^2=(\nabla m_x)^2+(\nabla m_y)^2+(\nabla m_z)^2$,
and 
$U$ (defined over all space) satisfies $\nabla\cdot(\nabla U+{\bf m})=0$.
The first term on the RHS of~(\ref{eq:H}) is the bending energy arising from spatial variations of the 
(now spatially dependent) magnetization ${\bf m}({\bf x})$, the second is the magnetostatic
energy, and the third is the energy due to coupling to an external magnetic field~${\bf h}$ (which is presumed
to be uniform). Crystalline anisotropy terms are neglected, given their negligibly small contribution; they can be easily included but will at most result in a
small modification of the much larger shape anisotropies arising from the magnetostatic term, to be discussed below.

Braun~\cite{Braun93} was the first to theoretically analyze thermal activation in the presence of
spatial variation of the magnetization density by considering magnetic reversal in
an infinitely long cylindrical magnet. However, Aharoni~\cite{Aharoni96}
pointed out that the energy functional used by Braun neglected important nonlocal magnetostatic 
energy contributions (the $\vert\nabla U\vert^2$ term in Eq.~(\ref{eq:H})), invalidating the result. 
Moreover, for submicron-scale magnets with large aspect ratio, finite system effects are also likely to play 
an important role; for example, simulations~\cite{BNR01,ERvE03} 
indicate that magnetization reversal in cylindrical-shaped particles
proceeds via propagation and coalescence of magnetic `end caps', nucleated at the cylinder
ends. Braun addressed these criticisms in~\cite{Braun99}, but both nonlocal effects and
nucleation and decay at the boundary presented an obstacle to further analytical work on these systems.

It turns out, however, that in quasi-2D systems these effects can be reduced under appropriate
physical conditions. Consider a thin film made of a soft magnetic material (e.g., permalloy), of any shape with diameter~$O(R)$ and thickness~$t$, and define
the dimensionless exchange length $\ell=\lambda/R$ and aspect ratio $k=t/R$, both much smaller than one and with $\ell^2\sim O(k\vert\log k\vert)$. Then an
asymptotic scaling analysis by Kohn and Slastikov~\cite{KS05a,KS05b} demonstrated that the problem simplifies from the full 3D situation; in particular, 
the nonlocal magnetostatic term in~(\ref{eq:H}) resolves into a local shape anisotropy with a strong energy penalty if the magnetization is not tangential to the
plane or its boundaries. Mathematically, if ${\bf\hat n}$ is the normal vector to the boundary at any point, then the Hamiltonian obtains additional boundary terms proportional to $({\bf m}\cdot{\bf\hat n})^2$; i.e., there is a large energy cost if the magnetization does not lie within the plane or is not tangential to its edges. (Nonlocal effects are still present but are an order or so magnitude smaller than the shape anisotropy terms.) The problem then becomes effectively two-dimensional and local, and for certain geometries is now amenable to analytical investigation.

In order to avoid nucleation at boundaries, Martens, Stein, and Kent~\cite{MSK05,MSK06} considered a thin annulus with inner radius~$R_1$, outer radius~$R_2$, and thickness~$t$, with the mean radius $R=(R_1+R_2)/2$ on the scale of $10^2$-$10^3$~nm and the thickness~$t$ of order~10~nm or less; with these parameters the authors were able to obtain analytic solutions for magnetization reversal times (in this case, from clockwise to counterclockwise magnetization orientation within the annular disk).  The magnetization reversal problem can now be mathematically modelled as a noise-induced transition in a Ginzburg-Landau scalar field theory perturbed by weak spatiotemporal noise.  

Using techniques developed for treating thermal or quantum nucleation of a state within a spatially extended system~\cite{Langer68,CC77,FJL82,MT95,MS01}, the authors found an experimentally realizable phase transition in the activation behavior, from Arrhenius to non-Arrhenius as either system size or external field (or both) are varied. For smaller ring sizes and/or larger fields reversal occurs through a rigid rotation of all the spins, as in the single domain case studied by N\'eel-Brown. In this regime activation shows the usual Arrhenius behavior. For larger sizes and/or smaller fields, reversal takes place through activation of an instanton-like saddle state~\cite{CC77}, where the magnetization configuration is described by Jacobi elliptic functions~\cite{AS65}. Here the transition is non-Arrhenius, due to the appearance of a zero mode corresponding to the rotational invariance of the instanton state. The phase diagram is shown in Fig.~1.

\begin{figure}
\label{fig:phase}
\centering \includegraphics[width=3in]{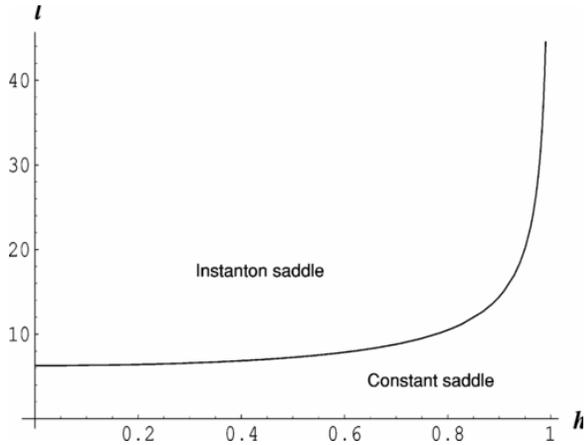} 
\caption{The phase boundary between the two activation regimes in the $(\ell,h)$-plane. In the shaded
region the instanton state is the saddle configuration; in the unshaded region, the constant state.  From~\cite{MSK06}.}
\end{figure}

Figs.~2 and 3 show the behavior of the activation energy $\Delta W$ as a function of annulus size and applied field.

\begin{figure}
\label{fig:W1}
\centering \includegraphics[width=3in]{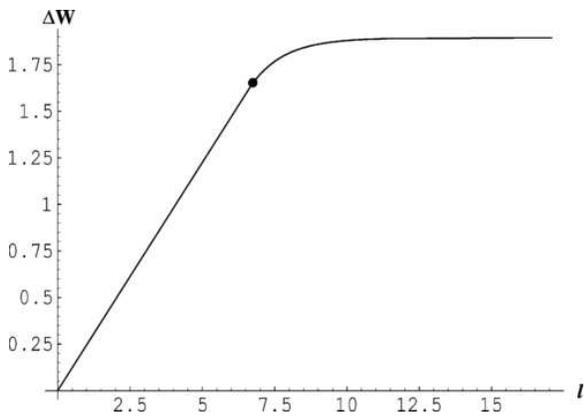} 
\caption{Activation energy $\Delta W$ for fixed $h = 0.3$ as $\ell$ varies. The dot indicates the transition from constant to instanton saddle configuration. From~\cite{MSK06}.}

\end{figure}
\begin{figure}
\label{fig:W2}
\centering \includegraphics[width=3in]{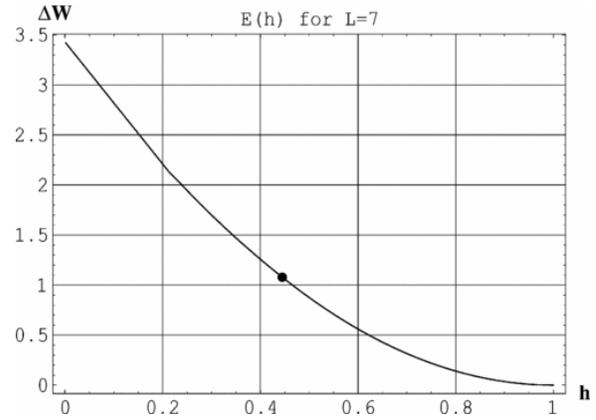} 
\caption{Activation energy $\Delta W$ for fixed $\ell=7$ as $h$ varies. The dot indicates the transition from instanton to constant saddle configuration. From~\cite{MSK06}.}
\end{figure}

Perhaps the most interesting part of the transition is its second-order phase transition nature, as shown by the divergence of the prefactor as the transition
from Arrhenius to non-Arrhenius behavior is approached from the Arrhenius side, as shown in Fig.~4.

\begin{figure}
\label{fig:prefactor}
\centering \includegraphics[width=3in]{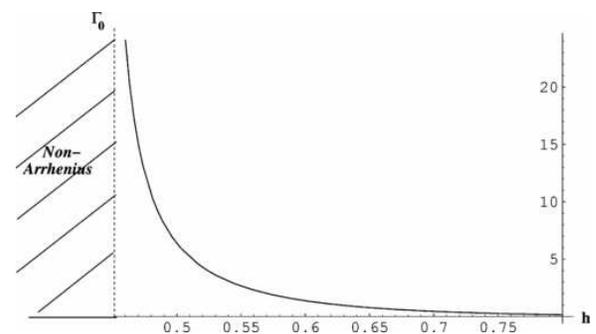} 
\caption{The prefactor $\Gamma_0$ (in units of $\tau_0$ vs.~$h$ when $\ell=7$ on the `constant saddle' side of the
transition. Here $\tau_0=\frac{m_0V(1+\alpha_2)}{\alpha\gamma E_0}$, where $V$ is the ring volume and $E_0$ is a characteristic energy scale depending on the various coupling constants in~(\ref{eq:H}) and the annulus parameters. The prefactor on the `instanton saddle' side of the transition acquires an additional temperature dependence, so doesn't appear on this graph. From~\cite{MSK06}.}
\end{figure}

Reversal rates at different temperatures, fields, and ring sizes are given in~\cite{MSK06} and we refer the interested reader there for more information.

\section{Numerical determination of transition states}
\label{sec:numerics}

The noise-induced transitions described in Sect.~\ref{sec:2D} are unusual in that their initial, final, and transition states can
all be solved for analytically.  In the vast majority of cases, at least one of these (usually the transition state), if not all three,
can only be found numerically. In this section we describe a family of computational techniques, known as chain-of-states methods, 
that can be adapted to many problems, both magnetic and non-magnetic, involving thermal activation over a barrier.

As already noted, in systems with magnetization dynamics satisfying $\mathbf{\dot{m}}=\mathbb{L}\mathbf{h}_{\mathrm{eff}}$
the stationary configurations are those in which $\frac{\delta\mathcal{E}}{\delta\mathbf{m}}=0$, and these are the
relevant configurations needed to compute the path in state space for thermally activated transitions. The
common ingredient of the numerical methods described here is that they compute a sequence of configurations (equivalently, a ``chain of states'')
that evolve toward the path of minimum overall energy that connects two (meta)stable configurations
$\mathbf{m}_{0}$ and $\mathbf{m}_{\mathrm{f}}$ in a continuous fashion~\cite{SimplifiedImprovedString2007,bessarabMethodFindingMechanism2015,jonssonNudgedElasticBand1998}.

In this section we briefly describe our implementation of one of these methods, known as the string
method, introduced by E, Ren, and vanden~Eijnden~\cite{ERvE03,kohn05,chaves11}.
We begin with its application to ferromagnetic annuli discussed in Sect.~\ref{sec:2D}, obtaining
a result that highlights the role of topologically
rigid textures as mediators for thermally activated transitions, which will be useful in Sect.~\ref{sec:skyrmions}.

The first step in a chain-of-states calculation is to identify two
distinct energy minima which are nearby (in the sense of a sequence of intermediate configurations)
in state space. To do this numerically it is often most efficient to use a standard micromagnetic simulator, many
of which use finite differences tools and are available in the public domain~\cite{donahueOOMMFUserGuide1999,vansteenkisteDesignVerificationMuMax32014}. 
There are multiple ways of carrying out this procedure: one easy alternative is to set the precessional
term~$\gamma({\bf m}\times {\bf h}_{\rm eff})$ to zero so that $\mathbf{\dot{m}}=\alpha\Bigl[\mathbf{m}\times(\mathbf{m}\times\mathbf{h}_{\mathrm{eff}})\Bigr]$, i.e., the dynamics is purely relaxational.
One can then choose an initial magnetization configuration and let it evolve toward the nearest local minimum;
by doing this multiple times one can then arrive at~$\mathbf{m}_{0}$ and~$\mathbf{m}_{\mathrm{f}}$. 

Once these states have been isolated, the chain-of-states method is used to find the saddle state between them.
In the string method, one creates an initial ``string'', i.e., a guessed sequence of configurations connecting $\mathbf{m}_{0}$ and $\mathbf{m}_{\mathrm{f}}$
in state space. Such a sequence will consist of $N+1$ space-discretized
configurations which we will call ``images''. For small systems with
a single domain, a simple global rotation between $\mathbf{m}_{0}$ and
$\mathbf{m}_{\mathrm{f}}$ along the length of the string constitutes
a good initialization for the string.  As systems grow in size and nonuniform magnetization configurations
become physically relevant, physical intuition becomes necessary to arrive at appropriate initial guesses 
of the string; the better the initial guess, the smaller the computation time. Ongoing work by a team that includes the
authors~\cite{Kuswikinprep} highlights the importance
of using the topological features of $\mathbf{m}_{0}$ and $\mathbf{m}_{\mathrm{f}}$
to assist in the generation of educated guesses of a string connecting them. Poor initial guesses 
result in (at best) greater computational times, and (at worst) convergence to the wrong string
connecting the two energy minima.

Once the initial guessed path has been chosen, the string is allowed to evolve
using a two-step iterative procedure~\cite{SimplifiedImprovedString2007}.
First, each image is allowed to evolve downward in energy by
a small amount; in our OOMMF-based~\footnote{This is an abbreviation for the Object Oriented MicroMagnetic Framework project at ITL/NIST~\cite{donahueOOMMFUserGuide1999}.} implementation~\cite{donahueOOMMFUserGuide1999} this is achieved
by running the purely relaxational simulation for a small amount of time.
Second, after all images have decreased in energy, an interpolation
is performed so that the arc-length distance between successive
images on the string is kept constant along the string. These two
steps are repeated until the string stops evolving.

At the end of the computation, the string converges to the minimum
energy path. The dependence on energy vs.~image number helps to identify
the energy minima and the saddle states. If $\mathbf{m}_{0}$ and
$\mathbf{m}_{\mathrm{f}}$ lie in neighboring basins of attraction,
they will be the only two energy minima. The image with the highest
energy value along the string is the saddle state.

To illustrate applications of this technique, we return to the
problem of the nanoring described in the previous section. Our numerical studies of annuli not only confirmed
the validity of the results presented in Sect.~\ref{sec:2D} but
revealed that, for sufficiently large rings, there exists a {\it multiplicity\/}
of possible metastable and transition states with distinct winding numbers, having energies separated in
discrete steps associated with the presence of $2\pi$ domain walls. 
The curvature of the annulus prevents degeneracies between two distinct 
transition states; they will have different energies depending on~(i) whether
they wind with or against the direction of the applied magnetic field (either clockwise or counterclockwise) 
and~(ii) their winding numbers (characterized by the number of $2\pi$ rotations the
magnetization configuration makes as one travels along the annulus for one revolution). 
Examples of two types of initial, final, and saddle states are shown in Fig.~5.
\begin{figure}[h]
\label{fig:annihilationof2piwall}
\includegraphics[width=3in]{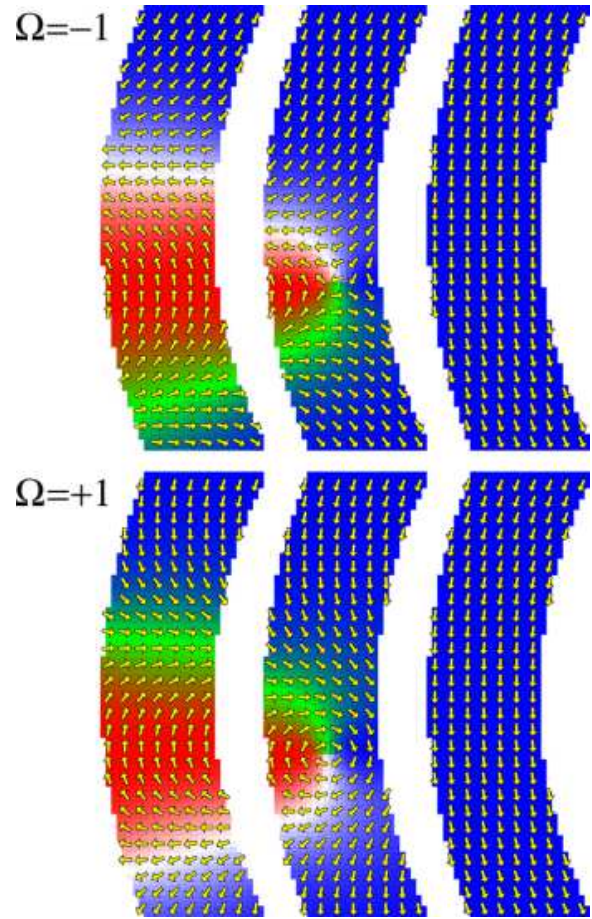}
\caption{Initial, saddle and final state for annihilation of two types of $2\pi$~domain walls in ferromagnetic nanorings. 
(Top) $2\pi$~domain wall with topological index~$\Omega=-1$.
(Bottom) $2\pi$ domain wall with topological index~$\Omega=+1$. From~\cite{chaves-oflynnStabilityPiDomain2010}.}
\end{figure}

Here the question we are interested in is the thermal stability of
these states, a problem suitable for analysis using the string method. 
The energies along the strings for the transitions of Fig.~5 are shown in Fig.~6.
\begin{figure}[h]
\label{fig:string2piannihilation}
\includegraphics[width=3in]{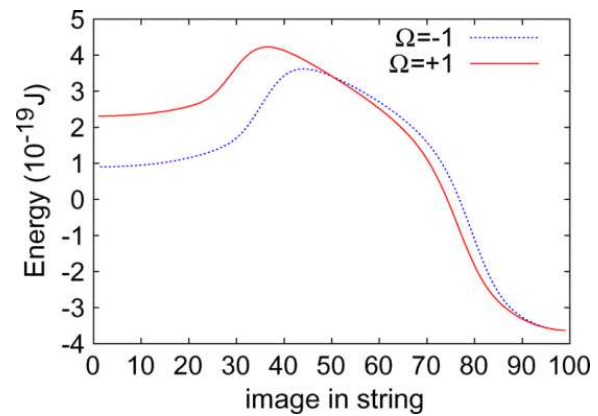}
\caption{Energy vs.~image number in a string method calculation for thermally
activated annihilation of $2\pi$ domain walls. From~\cite{chaves-oflynnStabilityPiDomain2010}.}
\end{figure}
In both cases, the annihilation process requires a topological defect
(vortex or antivortex) crossing the annular stripe. The saddle state
corresponds to the defect occupying the middle of the strip.

An important conclusion from this calculation is that the annihilation of
topologically protected objects (in this case a $2\pi$~wall) is mediated
by the motion of topological defects in the sample. Importantly, because 
$2\pi$ domain walls are topologically protected in bulk, these defects cannot arise in the interior but 
can only be created or destroyed at the in-plane edges. As a final observation,
the motion of these topological defects is amenable to treatment with
the use of a collective coordinate description of micromagnetic dynamics;
we return to this problem in Sect.~\ref{sec:skyrmions}.


\section{Thermal activation in magnetic textures I: Droplet solitons}
\label{sec:solitons}

In the previous two sections we considered thermally activated transitions between two globally reversed single-domain states, where the saddle state is either also single-domain or else a spatially-varying instanton state, depending on the system parameters and external conditions. Much research in this area, however, has focused on more complex configurations generally described as magnetic textures, which can be either topologically protected or not. In this section we discuss recent work on stochastic decay of (non-topological) droplet solitons, which have attracted a great deal of attention over the past few decades (for an introduction, see~\cite{KIK90}).

Magnetic droplet solitons are localized planar magnetic
textures that preserve their shape on timescales long
compared to typical magnon relaxation times~\cite{KIK90}; the spin configuration for a typical 
droplet soliton is shown in Fig.~7. 
\begin{figure}[h]
\label{fig:soliton}
\centering \includegraphics[width=3in]{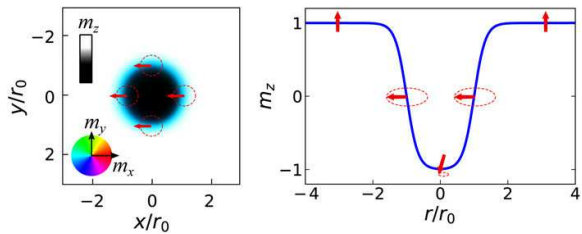} 
\caption{Schematic of a droplet soliton. Left: Spin configuration of a droplet
soliton of radius $r_0$ in a nanocontact. The color indicates the direction of magnetization.
Right: Profile through the droplet core. The spins precess about the crystalline anisotropy field, i.e., normal to the film
plane. At the droplet boundary, the precession amplitude is maximum, as
indicated by the red arrows. From~\cite{MK20}.}
\end{figure}

Unlike skyrmions, to be discussed in Sec.~\ref{sec:skyrmions}, droplet solitons are dynamical objects and are not topologically protected; 
without an external driving force, they 
collapse within a few precession cycles owing to dissipation. The usual method of stabilizing solitons is via a balance between the competing effects of 
dissipation and an energy input provided by an external current acting through spin transfer torque (STT)~\cite{Slonczewski1996,Berger1996}. STT is 
a method of transferring angular momentum from a spin-polarized current passing through a thick magnetic layer with fixed magnetization (called the ``fixed layer'') to magnetic moments residing within a thin magnetic layer (the ''free layer'', where the droplet soliton resides). If the system is driven by an STT-inducing current~$I$ with contact radius $\rho^{*}$, an additional nonconservative term must be added to~(\ref{eq:LLG}). This results in the Landau-Lifschitz-Gilbert-Slonczewksi~(LLGS) equation~\cite{bertotti06}: 
\begin{equation}
\mathbf{\dot{m}}=\mathbb{L}\mathbf{h_{\mathrm{eff}}}-\frac{\sigma\gamma m_0 V(\mathbf{r})}{1+\alpha^{2}}\left[\frac{\alpha{\bf m}\times{\bf\hat m}_p-{\bf m}\times{\bf m}\times{\bf\hat m}_p}{1+\nu{\bf m}\cdot{\bf\hat m}_p}\right]
\label{eq:LLGS}
\end{equation}
where ${\bf\hat m}_p$ is the magnetization direction of the fixed layer,
$\nu$ is a spin-torque anisotropy parameter (with $0\le\nu<1$),
$V(\mathbf{r})$ describes the spatial distribution of the (cylindrically
symmetric) current, and $\sigma$ is the reduced (i.e., dimensionless) current density~\cite{HSK10}.  

The effects of STT can be analyzed by introducing a term in the action corresponding to the energy density of a pseudopotential~\cite{CS20}
\begin{equation}
\label{eq:pseudo}
\mathcal{E}_{\mathrm{ST}}=\begin{cases}
\frac{m_0^2\sigma V(\rho)}{\alpha}\mathbf{m}\cdot{\bf \hat m}_p & \nu=0\\
\frac{m_0^2\sigma V(\rho)}{\alpha}\frac{\ln\left[1+\nu\mathbf{m}\cdot{\bf\hat m}_p\right]}{\nu} & \nu\ne0
\end{cases}
\end{equation}
with an associated spin torque effective field
\begin{equation}
\label{eq:stheff}
\mathbf{h}_{\mathrm{ST}}=-\frac{1}{m_0^{2}}\frac{\delta\mathcal{E}_{\mathrm{ST}}}{\delta\mathbf{m}}=-\frac{\sigma V(\rho)}{\alpha}\frac{{\bf\hat m}_p}{1+\nu{\bf m}\cdot{\bf\hat m}_p}\, .
\end{equation}

The total field is then $\mathbf{h}_{\mathrm{tot}}=\mathbf{h}_{\mathrm{eff}}+\mathbf{h}_{\mathrm{th}}+\mathbf{h}_{\mathrm{ST}}$, where $\mathbf{h}_{\mathrm{th}}$ is the thermal noise term. The magnetization dynamics including spin polarized currents and thermal fluctuations become:
\begin{eqnarray}
\label{eq:slls}
\dot{\bf m} &=&\mathbb{L}\mathbf{h_{\mathrm{tot}}}-\frac{m_0\sigma V(\rho)}{\alpha}\left[\frac{\mathbf{m}\times{\bf\hat m}_p}{1+\nu\mathbf{m}\cdot{\bf\hat m}_p}\right]\nonumber\\
 &=&\mathbb{L}\mathbf{h_{\mathrm{tot}}}-\frac{\gamma}{m_0}\nabla_{\mathbf{m}}\times\left(\mathbf{m}\mathcal{E}_{\mathrm{ST}}\right)
 \end{eqnarray} 
This is a compact version of the Stochastic-Landau-Lifzhitz-Slonczewski equation.

Eq.~(\ref{eq:slls}) makes explicit a further difficulty in applying tools of stochastic analysis to driven magnetic systems of this kind: the spin torque term has nonzero curl and therefore cannot be written as the gradient of a smooth potential (hence our use of a pseudopotential in~(\ref{eq:pseudo})). Moreover, a droplet soliton is an intrinsically dynamical object, stabilized by the rotation of the spins around the core, which adds an additional layer of difficulty in applying stochastic methods to soliton decay.

\subsection{Transformation to a rotating frame}
\label{subsec:rotating}

We address the second of these difficulties first. Eq.~(\ref{eq:slls}) describes the dynamics of the system in the fixed (laboratory) reference frame. Because the soliton rotates rigidly with a frequency $\omega_{\rm ST}=\sigma\gamma m_0V(\rho)/\alpha$, it is useful to transform to a coordinate system that rotates about ${\bf\hat m}_p$ with frequency $\omega_{\rm ST}$. In this frame the magnetization evolves as
\begin{equation}
\label{eq:rotating}
\dot{\tilde{\bf m}}_{\rm rotating\ frame}={\mathbb L}\mathbf{{\tilde h}_{\mathrm{tot}}}
\end{equation}
where a tilde denotes the transformed variable in the rotating frame.
In the system under investigation the energy functional is rotationally symmetric with respect
to ${\bf\hat m}_p$; as a consequence the fields in the rotating frame are time independent,
and (\ref{eq:rotating}) describes an autonomous dynamical system.
In this frame the magnetization evolves toward the nearest energy minimum~\cite{Serpico07}.
In the symmetric case when both ${\bf\hat m}_p$ and the external field are perpendicular to the plane of the thin film, the
pseudoenergy, derived in Appendix A of~\cite{CS20}, is given by
 \begin{equation}
 \label{eq:action}
 \mathcal{E}'_{\mathrm{tot}}=\vert\nabla'{\bf m}\vert^2-({\bf m}\cdot{\bf\hat z}_L)^2-{\bf m}\cdot{\bf h}'_0+\mathcal{E}'_{\rm ST}\, ,
\end{equation}
where all quantities are normalized to be dimensionless, primes refer to quantities in the rotating frame, ${\bf h}'_0$ is the external field, and ${\bf\hat z}_L$ is the unit vector in the $z$-direction in the laboratory frame and satisfies ${\bf\hat z}_L={\bf n}_\perp$, where ${\bf n}_\perp$ is the normal unit vector to the film plane. The first term on the RHS represents the contribution of the bending energy due to spatial variations in the magnetization, the second term is the magnetostatic energy which for two-dimensional systems results in a local shape anisotropy as discussed in Sec.~\ref{sec:2D}, and the third is the usual Zeeman term.

In the absence of noise, the total transformed energy in the rotating (primed) frame can  be shown to satisfy~\cite{CS20}
\begin{equation}
\label{eq:energy}
\frac{d{\mathcal E}_{\rm tot}'}{dt}=-\alpha\vert{\bf h}_{\rm tot}\vert^2+\Xi
\end{equation}
where
\begin{equation}
\label{eq:xi}
\Xi={\bf m}\cdot\Bigl(\frac{\delta{\mathcal E}_{\rm tot}'}{\delta{\bf m}}\times\nabla_{\bf m}{\mathbb E}'_{\rm ST}\Bigr)
\end{equation}
measures the degree of misalignment between fields
obtained from the total energy functional and the
spin torque pseudo-potential. More generally, it is a
measure of the time rate of energy change for a magnetization
configuration that appears stationary in the rotating frame.

When $\Xi=0$, i.e., the external field is in the same direction as~${\bf\hat m}_p$,
then as seen from~(\ref{eq:energy}) the energy monotonically decreases with time. When $\Xi\ne 0$, the magnitude
of the energy oscillates as the system rotates about ${\bf\hat m}_p$. If the energy change is zero
after one full orbit, i.e., if
\begin{equation}
\label{eq:selforbit}
\Delta{\mathcal E}_{\rm tot}'=\oint\Bigl(-\alpha\vert{\bf h}_{\rm tot}\vert^2+\Xi\Bigr) dt = 0
\end{equation}
then self-oscillations occur. Eq.~(\ref{eq:selforbit}) clarifies the meaning of $\Xi$:  it is
the spin-torque induced power density influx (the curl term in~(\ref{eq:slls})) which offsets
the energy loss due to the dissipative term $\alpha\vert{\bf h}_{\rm tot}\vert^2$, providing a 
steady-state soliton texture.

Therefore, by moving to the rotating frame, if external
laboratory fields are parallel to the fixed layer polarization then
there are steady-state solutions of the magnetization configuration; 
if not, then the solutions display small self-oscillations.
As we will see in the next section these solutions describe
either metastable configurations of the system or saddle states
(i.e., transition barriers).

\subsection{Extremal configurations of the action and soliton lifetimes}
\label{subsec:extremal}

With the appropriate laboratory setup~\cite{Kent18,MK20}, the driven dynamical system described above
can achieve a stable steady state in the absence of any noise~\cite{Mohseni13,HSK10}. However, the droplet
soliton is not topologically protected and in most parameter regimes is at best metastable, so in the presence of thermal noise 
it will eventually decay into a lower-energy spin wave state.

There are two decay mechanisms that have been theoretically explored. The first
decay mechanism~\cite{WIH16} is a linear instability at large currents.
In this region of parameter space the droplet soliton center drifts outside the nanocontact region, where spin-transfer torque is absent
and dissipation is therefore uncompensated, leading to a rapid decay of the soliton. This has been seen in
micromagnetic simulations~\cite{Lendinez15}.  The droplet soliton is nonetheless linearly stable
in other, experimentally accessible regions of parameter space, as shown in Fig.~2 of~\cite{WIH16}. 
Subsequent theoretical work~\cite{MH19}  demonstrated, however, that even in the linearly stable region of parameter space, a droplet soliton can be ejected
through thermal activation from the nanocontact region.

The second decay mechanism was investigated by the authors~\cite{CS20}, who studied thermal activation
of a stationary droplet soliton (i.e., one rotating about a fixed center in the laboratory frame). In this mechanism 
the center of the soliton remains stationary but thermal fluctuations over an energy barrier cause the droplet soliton
to decay to a spin wave state. This decay mechanism competes with the ejection mechanism; which one dominates droplet soliton decay will depend on
the experimental parameters. In the remainder of this section we discuss this second mechanism, and work throughout in
the rotating frame.

In most studies of activation employing the Kramers approach, the energy barrier $\Delta W$ in the resulting Arrhenius decay
$\tau_{\rm reversal}=\Gamma_0^{-1}\exp[\Delta W/k_BT]$ follows straightforwardly from a potential that describes the 
zero-noise deterministic dynamics.   However, in driven or otherwise nonequilibrium systems, especially those in which
detailed balance is not satisfied and no potential is available, it is more difficult to properly define a barrier.
In these cases, it is useful to employ a general path-integral approach to large deviations due to Wentzell and Freidlin~\cite{FW12},
which can be adapted to apply to nonequilbrium situations~\cite{MS83,GT86,HTB90,MS93,PKS16} such as the problem discussed here.

This approach requires determining the most probable path in state space between two locally stable configurations. 
The leading-order exponential term in the formula for $\tau_{\rm reversal}$ depends on the action difference $\Delta W$ 
between the starting metastable state and the transition state, which is usually but not always a saddle (i.e., hyperbolic fixed) 
point of the deterministic dynamics. These states are both extrema of the action functional, which in our case is given by~(\ref{eq:action}).
Given that $\vert{\bf m}\vert=m_0=1$ in reduced units, the magnetization can be parametrized in terms of fixed spherical coordinates
${\bf m}=(\sin\Theta\cos\Phi,\sin\Theta\sin\Phi,\cos\Theta)$. In the laboratory frame $\Phi$ is uniform in space and linearly dependent on time,
but is constant in both space and time in the rotating frame. The action extremum condition results in a nonlinear differential equation for
$\Theta$, which in the symmetric case under discussion is most naturally expressed in cylindrical coordinates. The extremal magnetization configurations satisfy
\begin{equation}
\label{eq:theta}
\dot\Theta_0=\frac{\partial^2\Theta_0}{\partial\rho'^2}+\frac{1}{\rho'}\frac{\partial\Theta_0}{\partial\rho}-\frac{1}{2}\sin(2\Theta_0)+\omega_h\sin\Theta_0=0
\end{equation}
where
\begin{equation}
\label{eq:omegah}
\omega_h=\Bigl[\frac{\sigma'V(\rho')}{\alpha}-h'_0\Bigr]
\end{equation}
where $h'_0$ is the magnitude of the external field (as in the preceding section, primes refer to quantities evaluated in the rotating frame). Eqs.~(\ref{eq:theta}) and~(\ref{eq:omegah}) were first derived in~\cite{HSK10} and were analyzed
in the case where the damping is a perturbation; we do not assume that here.

For intermediate nanocontact radii and low to intermediate currents the soliton profile, i.e.~the nonconstant solutions to Eqs.~(\ref{eq:theta}) and~(\ref{eq:omegah}), are given in Fig.~8.
\begin{figure}[h]
\label{fig:profiles}
\centering \includegraphics[width=3in]{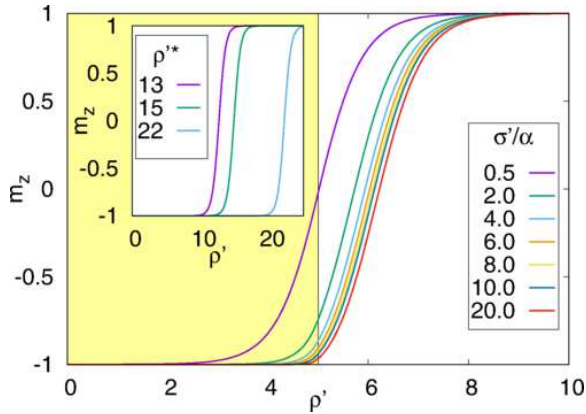} 
\caption{(Main figure) Stable droplet soliton profiles with asymmetry parameter $\nu=0$ for a variety of currents
and with nanocontact radius $\rho'^*= 5$. The nanocontact region is shaded yellow, from which
it can be seen that the transition region between the outside ($m_z=+1$) of the nanocontact and the soliton's inner core $(m_z=-1$)
occurs largely at the nanocontact's edge. (Inset) $m_z$ vs.~$\rho'$ at $\sigma'/\alpha = 0.13 $ for several
nanocontact radii.  From~\cite{CS20}.}
\end{figure}

With the boundary condition $m_z=+1$ as $\rho'\to\infty$, the uniform configuration $\Theta=0$ everywhere is the energetically stable state. 
However, if the current passes a certain threshold
$(\sigma'/\alpha)_{\rm crit}=(1+\nu)(h'_0+1)$, then the rotating droplet soliton is the energetically stable configuration. 
Inside the nanocontact region the magnetization switches to
$\Theta(\rho'=0)\to\pi$, with the magnetization configuration profile inside the nanocontact 
given by Fig.~8; in this parameter regime it is linearly stable against small displacements from the center of the
nanocontact. 

An important consequence of~(\ref{eq:theta}) is that, for a given applied current, there will be both (meta)stable and unstable
droplet soliton solutions, each corresponding to a specific frequency $\omega_0$ of rotation and with different values of $m_z$ at the origin; 
these are shown in Fig.~9 (see also Fig.~4 of~\cite{HSK10}). When $(\sigma'/\alpha)<(\sigma'/\alpha)_{\rm crit}$, as noted the 
solution $\Theta=0$ is stable. In addition there are two, physically relevant nonuniform solutions of~(\ref{eq:theta}).  One is a local energy minimum (metastable
droplet soliton), and the other is the saddle state (unstable droplet soliton). These correspond to the stable and unstable branches shown in Fig.~9. 

\begin{figure}[h]
\label{fig:omega}
\centering \includegraphics[width=3in]{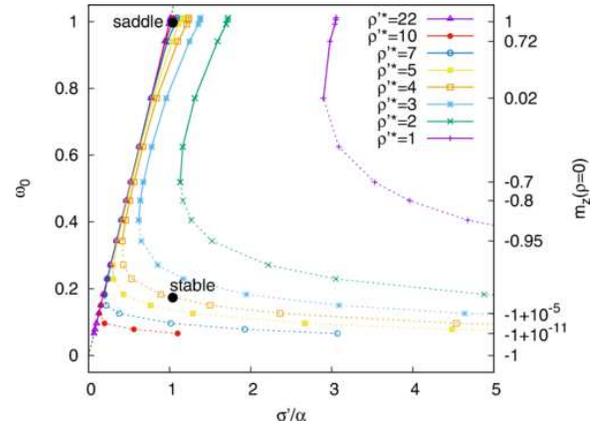} 
\caption{Droplet soliton solutions of~(\ref{eq:theta}) for different applied currents. Along the left axis the frequency for the infinite
nanocontact is shown; a dashed line of slope one depicts the infinite
nanocontact limit. It can be seen that already at moderate sizes the difference between
the solution at finite and infinite size is negligible.
All curves have two branches: the solid lines, where the
slope is positive, represent the saddle (unstable droplet soliton) states, and the dashed lines
indicate the corresponding stable droplet soliton states for the same nanocontact radius. The critical current for a
given nanocontact radius is the point at which the corresponding
curve crosses the uniform state, i.e., where $m_z(\rho'=0)=1$. The
sustaining current for each radius corresponds to the vertex where
the two branches meet. The filled black circles show saddle and stable droplet soliton
states for ${\rho'}^*=4.4$.  From~\cite{CS20}.}
\end{figure}

In~\cite{CS20} two kinds of numerical simulations were performed to confirm that
the solutions of~(\ref{eq:theta}) pictured in Fig.~8 indeed correspond to saddle configurations
in the energy landscape.  Both simulations model a magnetic material with exchange constant~$\lambda$
equal to 13~pJ/m and saturation magnetization $m_0 = 8\times 10^5$~A/m. The nanocontact diameter 
used is 50~nm ($\rho'^*=4.4$) and film thickness~1~nm. For these parameters the critical current is $\approx$~2.50~mA.

Results are shown in Fig.~10. The initial condition is the saddle
configuration corresponding to an unstable droplet soliton discussed above. Successive simulations
found two currents, $I^+$ and $I^-$, just above and below the critical current respectively; $\delta I=I^+-I^-=10^{-4}$.
The system evolves toward the uniformly magnetized state $m_z = 1$ for $I^-$ and toward the stable droplet soliton for $I^+$.

\begin{figure}[b]
\label{fig:overdamped}
\centering \includegraphics[width=3in]{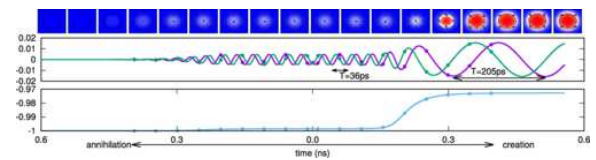} 
\caption{Magnetization configurations of the overdamped ($\alpha=1)$ dynamics, and the spatially averaged magnetization components vs.~time. Both simulations start with the same initial configuration, namely the unstable droplet soliton which is a saddle point between the uniform state (and the stable droplet soliton state. Time evolution occurs to the left for  
$I^-=2.491$~mA and to the right for $I^+= 2.492$~mA. Points in the curve are associated with the figures in the top row. Consistent with expectations, the low-amplitude droplet soliton (saddle state) has a higher frequency than the large-amplitude droplet soliton (stable droplet soliton). The precessional frequencies of the two configurations are $\omega_{\rm saddle} = 0.9964$ and $\omega_{\rm stable} = 0.1735$, both in units of $\frac{2\pi}{\gamma m_0 T}$. From~\cite{CS20}.}
\end{figure}

After simulating trajectories from saddle points the energy barriers were
measured. Results are shown in Fig.~11. For large values of
$\sigma/\alpha$ (Fig.~11a), the barrier for decay of a stable droplet soliton is close to linear
in the applied current, reflecting
the fact that once the nanocontact region is saturated the profile
remains unchanged; here $\mathbb{E}_{\rm ST}$ is roughly $\omega_{\rm ST}$ multiplied
by the nanocontact area. As $\sigma/\alpha$ is reduced (Fig.~11b), this
quasilinearity is lost for smaller radii.
\begin{figure}[h]
\label{fig:barriers}
\centering \includegraphics[width=3in]{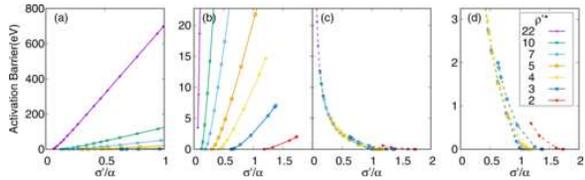} 
\caption{Energy barrier dependence on current in zero external field and with $\nu=0$.  (a,b) $\Delta^-$, the barrier for droplet soliton annihilation. (c,d) $\Delta^+$, the barrier for droplet soliton creation. As shown, the values of $\Delta^+$ are generally much smaller than those for $\Delta^-$. To better illustrate the range of currents for which the barriers $\Delta^+$ and 
$\Delta^-$, (b) and (c) are juxtaposed with the same axis scale. From~\cite{CS20}.}
\end{figure}
The creation barrier $\Delta^+$ (Figs.~11c and 11d) has a singularity
at $\sigma/\alpha = 0$ and decreases rapidly toward the critical
current, consistent with expectations. As the nanocontact radius is reduced, 
the region of bistability is reduced and the barriers for droplet soliton creation grow.

\section{General theory of noise-induced transitions in the presence of arbitrary dynamics}
\label{subsec:generalization}

Before proceeding to a discussion of the noise-induced creation and annihilation of topologically protected skyrmions, 
we present a more general formalism designed to analyze the problem of transition rates for textures supported by arbitrary dynamics.  
To do so, we add an arbitrary torque $\bm{\tau}$ to the equation of motion: 
\begin{equation}
\label{eq:implicit}
\frac{\partial\mathbf{m}}{\partial t}=-\gamma\left(\mathbf{m}\times\mathbf{h}_{\mathrm{eff}}\right)+\alpha\left(\mathbf{m}\times\frac{\partial\mathbf{m}}{\partial t}\right)+\bm{\tau}\, .
\end{equation}
Because the magnitude of the magnetization is constant, all terms
in the above equation are tangent to the unit sphere. Taking the inner
product of both sides with $\frac{\partial\mathbf{m}}{\partial t}$, we then obtain
\begin{align}
\frac{\partial\mathbf{m}}{\partial t} & =\mathbb{L}\left(\mathbf{h}_{\mathrm{eff}}+\mathbf{m}\times\frac{\bm{\tau}}{\gamma}\right)\label{eq:compactLLGgeneraltorque}
\end{align}

Here it is useful to use the Helmholtz-Hodge decomposition \citep{fan18},
which states that a vector field tangent to the unit sphere can always be
decomposed as a sum of a divergence-free term and a curl-free
term. The additional arbitrary torque can then be written as
\begin{equation}
\bm{\tau}=\left[\nabla_{\mathbf{m}}W+\nabla_{\mathbf{m}}\times\mathbf{A}\right],\label{eq:extratorque}
\end{equation}
where the vector $\mathbf{A}$ can be chosen parallel to the unit magnetization vector, 
i.e., $\mathbf{A}=A_{m}\left(\theta,\phi\right)\mathbf{\hat{m}}$.
Taking the divergence and curl of Eq.~(\ref{eq:extratorque})
leads to two decoupled equations that allow us  to find both $\mathbf{A}$ annd the field potentials $W$:
\begin{equation}
\nabla\cdot\bm{\tau}=\nabla^{2}W
\end{equation}
\begin{equation}
\mathbf{m}\cdot\left(\nabla\times\bm{\tau}\right)=\nabla^{2}A_{m}.
\end{equation}

After finding $A_{m}$ and $W$, we can define a pseudo-potential
and an associated field:
\begin{align}
\mathcal{E_{\tau}} & =-\frac{1}{\gamma}\left(A_{m}+\frac{W}{\alpha}\right)\\
\mathbf{h}_{\tau} & =\nabla_{\mathbf{m}}\frac{1}{\gamma}\left(A_{m}+\frac{W}{\alpha}\right)
\end{align}
so that the total deterministic field is the sum of the associated field $\mathbf{h}_{\tau}$ and the
variational derivative of the original energy functional:
\begin{equation}
\mathcal{E}_{\mathrm{det}}=\mathcal{E}_{\mathrm{eff}}+\mathcal{E}_{\tau}\qquad\mathbf{h}_{\mathrm{det}}=\mathbf{h}_{\mathrm{eff}}+\mathbf{h}_{\tau}\, .
\end{equation}

In order to recover the original magnetization dynamics an extra
term must be added to the equation of motion, which now becomes: 
\begin{align}
\frac{\partial\mathbf{m}}{\partial t} & =\mathbb{L}\mathbf{h}_{\mathrm{det}}-\nabla_{\mathbf{m}}\times\left[\frac{\mathbf{m}W}{\alpha}\right]=\mathbb{L}\mathbf{h}_{\mathrm{det}}+\mathbf{m}\times\nabla_{\mathbf{m}}\left[\frac{W}{\alpha}\right].\label{eq:explicit}
\end{align}

It is enlightening to compare the roles that the potentials $\mathbf{A}$
and $W$ play in the two different versions of the equation of motion of the magnetization.
The divergence free part of $\bm{\tau}$ appears in the ``implicit'' version~(\ref{eq:implicit}) as a curl term, 
but in the ``explicit'' version~(\ref{eq:explicit}) it appears as a typical field term.
In contrast, the curl-free term in the implicit equation requires the introduction of the additional curl term in the
explicit version. This distinction becomes important given that
Brown's theory of thermally-induced reversal is applicable for magnetization dynamics of the form
\begin{eqnarray}
\label{eq:Brown}
\frac{\partial\mathbf{m}}{\partial t}&=&\mathbb{L}\left(-\frac{\delta\mathcal{E}}{\delta\mathbf{m}}+\mathbf{h}_{\mathrm{therm}}\right)\nonumber\\
\mathbf{h}_{\mathrm{therm}}&=&\sqrt{2\eta}\mathbf{\dot{W}}
\end{eqnarray}
with a thermal field $\mathbf{h}_\mathrm{therm}$ of strength
$\eta$ and Gaussian-distributed white noise $\dot{\mathbf{W}}$. This system
satisfies several important features shared by systems in which Kramers'
theory of escape rates is applicable:

\begin{enumerate}
\item Critical points of the energy landscape (i.e., magnetization configurations that satisfy $\frac{\delta\mathcal{E}}{\delta\mathbf{m}}=0$) are
necessarily critical dynamical points (where $\frac{d\mathbf{m}}{dt}=0$).
\item The Boltzmann distribution $P\left(\mathbf{m}\right)\propto e^{-\frac{E(\mathbf{m})}{k_{\mathrm{B}}T}}$
is a stationary solution of the associated Fokker-Planck equation.
\item From the stationarity of the Boltzmann distribution, the fluctuation
dissipation theorem provides a relation between temperature, damping
and the strength of the noise terms $\eta\propto\alpha k_{\mathrm{B}}T$.
\end{enumerate}

As noted in Sect.~\ref{sec:solitons}, if rotational symmetry
is present the last term in~(\ref{eq:explicit}) can be absorbed
by a change of reference frame; if that fails, there is no guarantee that the system will
reach a Boltzmann distribution in the steady state (if one exists).

Generalizing the previous result for the time rate of energy change we find
\begin{align}
\frac{d\mathcal{E}}{dt} & =-\alpha\mathbf{h}_{\mathrm{det}}^{2}+\Xi\\
\Xi & =\mathbf{m}\cdot\left(\frac{\delta\mathcal{E}_{\mathrm{det}}}{\delta\mathbf{m}}\times\nabla_{\mathbf{m}}\left[\frac{W}{\alpha}\right]\right).
\end{align}

We conclude that if the magnetization dynamics cannot be reduced to the form~(\ref{eq:Brown}),
then the usual formalism used in  Kramer's theory of escape rates may not be applicable and the stochastic
dynamics must be studied with more sophisticated tools such as the Wentzell-Freidlin
theory~\cite{FW12}, which was used in the presence of spin-tranfer torque for a macrospin~\citep{kohn05,chaves11}, i.e.,
a system with uniform magnetization. 

\section{Thermal activation in magnetic textures II: Skyrmions}
\label{sec:skyrmions}

Skyrmions are nonsingular, topologically stable, finite-energy particle-like configurations, introduced in 1961 
as stable baryonic excitations within a class of nonlinear $\sigma$-models that were used to investigate the 
structure of the nucleon~\cite{Skyrme61}. Since then they have found several applications in condensed-matter
physics, including chiral nematic liquid crystals~\cite{FZ11}, Bose-Einstein condensates~\cite{KH01}, and thin magnetic films,
particularly in the context of spintronics~\cite{FNC17,BLB18}, where their particle-like behavior is potentially useful for
high-density data storage applications. These magnetic skyrmions will be the focus of this section.

\subsection{Fundamentals}
\label{subsec:funds}

As noted above, magnetic skyrmions are magnetization textures that are topologically protected, i.e., cannot be continuously
deformed (strictly speaking, in the continuum limit) to the uniform state. Topological textures in general represent a continuous mapping from order parameter space to physical space.
In the case of interest here, the magnetization vectors live on a planar surface with the boundary condition that all spins far from the origin point 
in a single direction (say up).  Because each magnetization vector has unit magnitude and can point in any direction in $\mathbb{R}^3$, its value at any point in space can be
described as a point on $S^2$, where $S^2$ is the two-sphere. Consequently, an arbitrary magnetization configuration (again, in the continuum limit) on the plane with the boundary condition~$m_z=+1$ at infinity can be mathematically described as a mapping from $S^2\to S^2$. 

In algebraic topology all such mappings can be sorted into homotopy classes, with each class consisting of the set of all configurations continuously deformable into one another.
These classes form a group, and in the case of continuous mappings from $S^2\to S^2$, this group is isomorphic to the integers~$\mathbb{Z}$~\cite{Armstrong79}.
The integer associated with a particular magnetization configuration corresponds to its winding number~$\Omega$ about $S^2$, defined as
\begin{equation}
\label{eq:winding}
\Omega=\frac{1}{4\pi}\int_{\mathbb{R}^2}{\bf m}\cdot(\partial_x{\bf m}\times\partial_y{\bf m})\ d^2x\, .
\end{equation}
The uniform configuration and all configurations continuously deformable to it have $\Omega=0$; we will be interested in configurations with winding number $\Omega=\pm 1$,
also known as magnetic skyrmions.  Two examples of skyrmions with unit winding number are shown in Fig.~12.
\begin{figure}[h]
\label{fig:skyrmions}
\centering \includegraphics[width=3in]{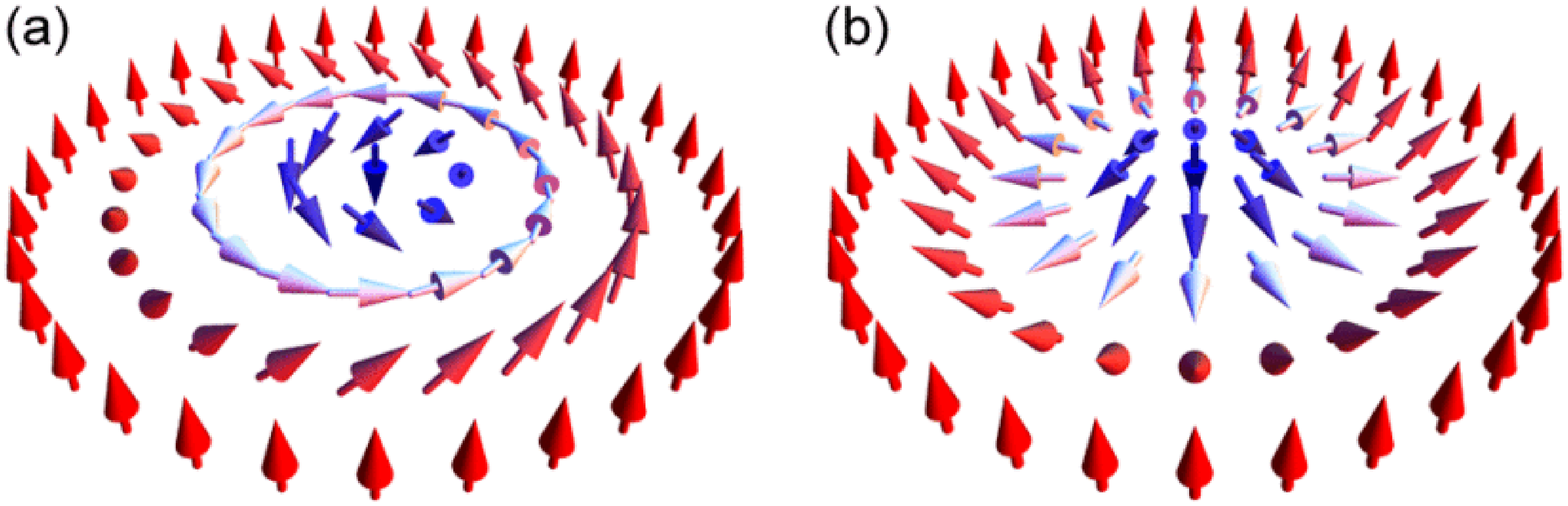} 
\caption{Two types of skyrmions with unit winding number. (a) Bloch-type
skyrmion, in which the magnetization direction rotates
perpendicular to the radial direction  when moving from the
core to the periphery. (b) Neel-type skyrmion,  in which the spins
rotate within the radial planes from the core to the periphery.  From~\cite{KHZZZ16}.}
\end{figure}

Given that skyrmions are chiral structures, one looks for them in systems with strong chiral interactions.  Specifically, they arise in non-centrosymmetric lattices, and are stabilized by Dzyaloshinskii-Moriya interactions~(DMI) arising from strong spin-orbit couplings induced by the lack of inversion symmetry.  The DMI between two neighboring magnetic moment ${\bf m}_i$ and ${\bf m}_j$ 
is described by the interaction Hamiltonian
\begin{equation}
\label{eq:DMI}
H^{\rm DMI}_{ij}={\bf D}_{ij}\cdot({\bf m}_i\times{\bf m}_j)\, ,
\end{equation}
where the Dzyaloshinskii-Moriya vector ${\bf D}_{ij}$ is determined by the lattice. Early work on magnetic skyrmions used ultrathin magnetic films
(e.g., Fe monolayers or PdFe bilayers) epitaxially grown on a heavy metal such as Ir~\cite{FNC17}; in these systems inversion symmetry is broken 
at the interface and the heavy metal provides strong spin-orbit coupling.

The combination of topological protection against decay and the particle-like or solitonic nature of skyrmions makes
them prime candidates for storing, transferring, and manipulating information within magnetic thin films. However, as
discussed at the end of~Sect.~\ref{sec:numerics}, structures which are topologically protected in bulk remain susceptible to decay
at the film boundaries, and it is therefore important to understand thermally-activated decay of skyrmions due to the creation of 
textures at the edges that can ``unwind'' the skyrmion.  This is the subject of ongoing work~\cite{Kuswikinprep}, which we briefly report on below.
(An important side note:  topological protection strictly applies only in the continuum limit of an infinite system, and as briefly noted in Sect.~\ref{subsec:applications}, in physical systems this protection can be lost due either to the discrete nature of the lattice, leading to another form of decay in which the skyrmion shrinks to a few lattice spacings and disappears, or to finite-size effects, in which the soliton expands beyond the system size.) 

We begin by discussing a general method that is useful not only for problems in magnetism but for many other systems describable by
classical or quantum field theories.

\subsection{Collective coordinate methods in dynamical field theories}
\label{subsec:collective}

In previous sections we have seen that fairly complicated magnetization textures 
often constitute transition states, even between uniform or other simple ground state configurations.
In more complex situations, including driven systems, such textures may themselves constitute the stable or metastable states.  
In these situations one must usually work in an infinite-dimensional
state space; however, there exist techniques to simplify the analysis by reducing the description
of the problem to just a few relevant collective coordinates. This is particularly useful in analyzing transitions
between topologically-protected textures such as skyrmions and other configurations.

The general idea is to reduce the full, infinite-dimensional 
magnetization field to a small number of coordinates that
incorporate only the softest modes, which dominate the long-time dynamics --- and
therefore reflect most closely the key features of the energy landscape~\citep{clarkeDynamicsVortexDomain2008,tretiakovDynamicsDomainWalls2008}.
The earliest demonstration of this approach was introduced by Thiele 
in his work on the motion of domain walls \citep{thieleSteadyStateMotionMagnetic1973} in magnetic systems.
Although he considered a situation where the collective coordinates
represent the position of a domain wall, the collective coordinates
need not represent positions in space: in principle (and often in practice), any parameter
used to describe a magnetization profile can serve as a collective coordinate.
For example, in studies of conservative droplet solitons it is convenient
to use frequencies and phases as the collective coordinates~\citep{bookmanAnalyticalTheoryModulated2013,bookmanPerturbationTheoryPropagating2015};
in this case, the equations of motion then have a Hamiltonian structure
\citep{mckeeverCharacterizingBreathingDynamics2019}. 

The reduction to collective coordinates starts by selecting magnetization
configurations with time-dependent collective coordinates~$\mathbf{q}(t)=\left\{ q_{s}\right\} $
as parameters for the spatial profiles
\begin{equation}
\mathbf{m}(\mathbf{r},t)=\mathbf{m}_{0}(\mathbf{r};\mathbf{q}(t))\, .
\end{equation}
Purely as a notational convenience, one can define $q_{0}=t$ 
so that the rate of change of magnetization becomes
\begin{equation}
\dot{\mathbf{m}}_{0}=\sum_{i}\frac{\partial\mathbf{m}_{0}}{\partial q_{i}}\dot{q_{i}}\, .
\end{equation}

The energy density $\mathcal{E}$ at a point $\mathbf{r}$ in the sample depends on $\mathbf{m}_{0}$ and its spatial derivatives:
\begin{equation}
\mathcal{E}=\mathcal{E}\left(\mathbf{m}_{0}\left(\mathbf{r};\mathbf{q}\right),\frac{\partial\mathbf{m}_{0}\left(\mathbf{r};\mathbf{q}\right)}{\partial r_{i}};\mathbf{r},\mathbf{q}\right)\, .
\end{equation}
Using this we can calculate the {\it total\/} force acting on the magnetization texture:
\begin{align}
F_{s} & =-\int\text{\ensuremath{\frac{\ensuremath{d\mathcal{E}}}{dq_{s}}}}d^{3}\mathbf{r}\label{eq:generalforce}\\
 & =-\int\frac{\delta\mathcal{E}}{\delta\mathbf{m}_{0}}\cdot\text{\ensuremath{\frac{\ensuremath{d}\mathbf{m}_{0}}{dq_{s}}}}d^{3}\mathbf{r}-\oint\mathbf{P}_{j}\cdot\mathbf{\hat{n}}_{j}d^{2}\mathbf{r}\, .
\end{align}
The second term on the RHS accounts for an effective pressure at the boundary
of the magnetic body, and is dropped for infinite systems.
For finite systems, the functions~$\mathbf{m}_{0}$ that define the profiles  are customarily chosen so that
\begin{equation}
\mathbf{P}_{j}=\left[\frac{\partial\mathcal{E}}{\partial\left(\partial_{j}\mathbf{m}_{0}\right)}\cdot\frac{\ensuremath{d}\mathbf{m}_{0}}{dq_{s}}\right]=0
\end{equation}
at the boundaries of the system. Under this condition the total force on the magnetization texture reduces to 
\begin{equation}
F_{s}\equiv-\int\frac{\delta\mathcal{E}}{\delta\mathbf{m}_{0}}\cdot\text{\ensuremath{\frac{\ensuremath{d}\mathbf{m}_{0}}{dq_{s}}}}d^{3}\mathbf{r}\label{eq:generalizedforce}\, .
\end{equation}
We now need an expression for $\frac{\delta\mathcal{E}}{\delta\mathbf{m}_{0}}$
which we obtain by inverting the equation of motion (\ref{eq:explicit}), including a thermal field. 
Because $\mathbf{\dot{m}}\perp\mathbf{m}$, we can define $\mathbb{L}^{-1}$ in the subspace normal to $\mathbf{m}$,
even though in three dimensions one of the eigenvalues of $\mathbb{L}$
is zero (for the sub-space parallel to $\mathbf{m}$). Using the identity matrix $\mathbf{1}$ this inverse can be written as
\begin{align}
\mathbb{L}^{-1} & =-\frac{1}{\gamma_{0}}\left[\mathbf{m}\times+\alpha\mathbf{m}\times\mathbf{m}\times\right]\\
 & =\frac{1}{\gamma_{0}}\left[\alpha\mathbf{1}-\mathbf{m}\times\right].
\end{align}
For convenience, we also use the relation
\begin{equation}
\mathbb{L}^{-1}\mathbf{m}\times=\frac{1}{\gamma_{0}}\left[\mathbf{1}+\alpha\mathbf{m}\times\right]\, .
\end{equation}
The inverse equation of motion~(\ref{eq:explicit}) then provides the expression we are looking for:
\begin{align}
\frac{\delta\mathcal{E}}{\delta\mathbf{m}_{0}}= & \mathbf{h}_{\mathrm{therm}}-\mathbb{L}^{-1}\mathbf{m}\times\nabla_{\mathbf{m}}\left[\frac{W}{\alpha}\right]-\mathbb{L}^{-1}\dot{\mathbf{m}}\label{eq:inverseLLG}\\
= & \mathbf{h}_{\mathrm{th}}+\frac{1}{\gamma_{0}}\left[\mathbf{1}+\alpha\mathbf{m}_{0}\times\right]\nabla_{\mathbf{m}}\left[\frac{W}{\alpha}\right]\nonumber \\
 & -\sum_{t}\frac{1}{\gamma_{0}}\left[\alpha\mathbf{1}-\mathbf{m}_{0}\times\right]\frac{\partial\mathbf{m}_{0}}{\partial q_{t}}\cdot\dot{q_{t}}\, .
\end{align}

Substituting this into the definition of the generalized force~(\ref{eq:generalizedforce}), we obtain an integral form of the equation of motion:
\begin{align}
F_{s}+T_{\mathrm{s}}+W_{s}+B_{s} & =\sum_{t}\left(\Gamma_{st}-G_{st}\right)\dot{q}_{t}\label{eq:generalizedthiele}
\end{align}
with the following definitions for each of the terms: 
\begin{align}
F_{s}= & -\int\frac{\delta\mathcal{E}}{\delta\mathbf{m}_{0}}\cdot\text{\ensuremath{\frac{\ensuremath{d}\mathbf{m}_{0}}{dq_{s}}}}d^{3}\mathbf{r}\\
T_{\mathrm{s}}= & \int\left[\mathbf{h}_{\mathrm{th}}\cdot\text{\ensuremath{\frac{\ensuremath{d}\mathbf{m}_{0}}{dq_{s}}}}\right]d^{3}\mathbf{r}\\
W_{s}= & \frac{1}{\gamma_{0}}\int\left[\text{\ensuremath{\frac{\ensuremath{d}\mathbf{m}_{0}}{dq_{s}}}\ensuremath{\cdot}}\nabla_{\mathbf{m}}\left[\frac{W}{\alpha}\right]\right]d^{3}\mathbf{r}\\
B_{s}= & \frac{\alpha}{\gamma_{0}}\int\left[\text{\ensuremath{\mathbf{m}_{0}}\ensuremath{\cdot}}\left[\nabla_{\mathbf{m}}\left[\frac{W}{\alpha}\right]\times\frac{\ensuremath{d}\mathbf{m}_{0}}{dq_{s}}\right]\right]d^{3}\mathbf{r}\\
\Gamma_{st}= & \frac{\alpha}{\gamma_{0}}\sum_{t}\int\left[\text{\ensuremath{\frac{\ensuremath{d}\mathbf{m}_{0}}{dq_{s}}}}\cdot\frac{\partial\mathbf{m}_{0}}{\partial q_{t}}\dot{q_{t}}\right]d^{3}\mathbf{r}\\
G_{st}= & \frac{1}{\gamma_{0}}\sum_{t}\int\left[\mathbf{m}_{0}\cdot\left[\frac{\partial\mathbf{m}_{0}}{\partial q_{t}}\times\text{\ensuremath{\frac{\ensuremath{d}\mathbf{m}_{0}}{dq_{s}}}}\right]\dot{q_{t}}\right]d^{3}\mathbf{r}
\end{align}
The deterministic version of~( \ref{eq:generalizedthiele}) is identical
to that of Thiele under the restrictions $W=B=0$. 


For the final step the noise covariance matrix must be constructed for the collective
coordinates~\citep{kamppeterStochasticVortexDynamics1999}:
\begin{equation}
\left\langle T_{\mathrm{s}}T_{t}\right\rangle =\left\langle \iint\left[\mathbf{h}'_{\mathrm{th}}\cdot\text{\ensuremath{\frac{\ensuremath{d}\mathbf{m}'_{0}}{dq_{s}}}}\right]\left[\mathbf{h}_{\mathrm{th}}\cdot\text{\ensuremath{\frac{\ensuremath{d}\mathbf{m}_{0}}{dq_{t}}}}\right]d^{3}\mathbf{r}d^{3}\mathbf{r}'\right\rangle\, .
\end{equation}

In the limit of small noise, the magnetization terms can be taken out of the average over noise realizations~\citep{miltatBrownianMotionMagnetic2018} to obtain
\begin{equation}
\label{eq:covariantmatrix}
\left\langle T_{\mathrm{s}}(t)T_{t}(t')\right\rangle =\delta(t-t')\int\left[\text{\ensuremath{\frac{\ensuremath{d}\mathbf{m}{}_{0}}{dq_{s}}}}\cdot\text{\ensuremath{\frac{\ensuremath{d}\mathbf{m}_{0}}{dq_{t}}}}\right]d^{3}\mathbf{r}'\, .
\end{equation}
The expectations $\left\langle T_{\mathrm{s}}\right\rangle$ of the noise terms can be taken to be zero. 

This completes the description of the steps that are generally needed to reduce the stochastic Landau-Lifschitz-Gilbert
equation to a low-dimensional system. Naturally, the validity of
low-dimensional reductions depends strongly on appropiate choices
of collective coordinates $\mathbf{q}$ and profiles $\mathbf{m}_{0}(\mathbf{r},\mathbf{q})$.
Once the reduction has been achieved, the results of low-dimensional large
deviation theory can be used to make further predictions. Not surprisingly, the major challenge to fully incorporate stochastic
processes within a reduced-dimensional model is usually the calculation of the covariant matrix~(\ref{eq:covariantmatrix}). 

\subsection{Applications}
\label{subsec:applications}

We conclude this section by pointing to recent work that exemplifies
the use of collective coordinates for the study of thermally activated
transitions between micromagnetic states. Some of these studies include
explicit derivations of the corresponding covariant matrices.

As mentioned in Sect.~\ref{sec:solitons}, random perturbations of
magnetic droplet solitons have been studied using a Hamiltonian framework~\citep{bookmanAnalyticalTheoryModulated2013,mooreStochasticEjectionNanocontact2019,willsDeterministicDriftInstability2016,bookmanPerturbationTheoryPropagating2015}.
Because the soliton is a dynamic texture, as already noted the frequency
of precession and the center and velocity of the droplet are the natural
collective coordinates to use for determining the stochastic decay
rate of the soliton.

A separate study of magnetization reversal on magnetic disks, based
on the stochastic motion of a domain wall \citep{bouquinSpintorqueInducedWall2021,bouquinStochasticProcessesMagnetization2021},
uses the wall position and the wall's tilt (i.e., the in-plane orientation
at the center wall) to analyze magnetization reversal. This work compares
results from micromagnetic simulations at room temperature to key
features of the deterministic, reduced-dimensional model, and is able
to predict whether the reversal occurs with a single sweep of the
domain wall, or whether the wall oscillates as it crosses the center
of the disk. A similar model~\citep{statutoMicromagneticInstabilitiesSpintransfer2021}
identifies the basins of attraction of two stable configurations in
the presence of spin currents.

Returning now to magnetic skyrmions, a reduced model of skyrmion motion~\citep{bernand-mantelMicromagneticTheorySkyrmion2021,bernand-mantelUnravelingRoleDipolar2020},
based on Belavin-Polyakov profiles~\citep{belavinMetastableStatesTwodimensional1975},
has been used to obtain skyrmion lifetimes for an annihilation path
that consists of the skyrmion shrinking to zero radius; here, the
annihilation process depends on the discretization of the magnetization
configuration. Another study compares the barrier for a skyrmion shrinking
to zero radius to one for skyrmions growing beyond the boundaries
of the magnetic material, again effectively destroying themselves~\citep{riverosFieldDependentEnergyBarriers2021}.

Our ongoing study of ferromagnetic nanodisks examines the influence
of the interfacial Dzyaloshinskii-Moriya interaction on their energy
landscapes. In these systems the DMI strength can be described with
a single scalar $D$. As $D$ changes, the nature of both the energy minima
and the transition states changes qualitatively,  requiring
different choices of collective coordinates. 

\begin{figure*}
\includegraphics[width=6in]{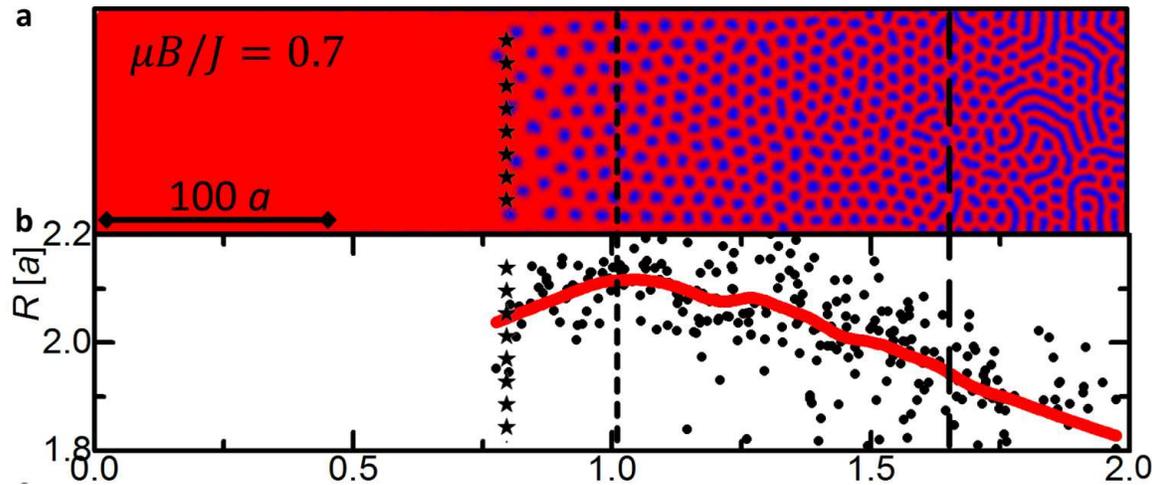}\caption{\label{fig:phasesbyD} (a) Equilibrium magnetization configurations
of a film subjected to a perpendicular external field, as $D/D_{c}$ varies; red and blue represent upward and downward magnetizations,
respectively. (b) Skyrmion radii $R$ (in units of the discrete lattice spacing~$a$) as a function of $D$.  From~\citep{siemensMinimalRadiusMagnetic2016}.}
\end{figure*}
We start by presenting a celebrated phase diagram~\citep{siemensMinimalRadiusMagnetic2016,fertMagneticSkyrmionsAdvances2017},
summarized in Fig.~\ref{fig:phasesbyD}, which encapsulates the effect of $D$ on the magnetic phase of chiral
magnets, for a system in which the gradient of $D$ lies along $x$. 
As an aid to its interpretation, we note that domain walls separating regions of opposite
magnetization have a characteristic energy density per unit length~$\lambda_{w}=\left(4\sqrt{AK_{\mathrm{eff}}}-\pi\left|D\right|\right)$. 
It is easily seen that larger~$D$ favors the growth of domain walls. From~(\ref{eq:DMI}) it is apparent that the sign of $D$ determines the sense of rotation
of the magnetization. It is convenient to define a critical value,
$D_{c}=\frac{4}{\pi}\sqrt{AK_{\mathrm{eff}}}$, which indicates the
point at which $\lambda_{w}=0$.

For $D<D_{c}$, the uniform magnetization is the lowest-energy state;
here the exchange interaction, which penalizes deviations from
uniformity, dominates the DMI which favors chirality. For $D>D_{c}$, 
the minimum-energy configuration is chiral~\citep{leonovPropertiesIsolatedChiral2016}: the magnetization
rotates about ${\bf D}_{ij}$. Depending on whether ${\bf D}_{ij}$
is parallel or perpendicular to the vector $\mathbf{r}_{ij}$ pointing
from~$\mathbf{m}_{i}$ to~$\mathbf{m}_{j}$, 
the profile $\mathbf{m}(x)$ can be either helical (${\bf D}_{ij}\parallel\mathbf{r}_{ij}$)
or cycloidal (${\bf D}_{ij}\perp\mathbf{r}_{ij}$). The helical period
$L_{\mathrm{hel}}=\frac{4\pi A}{D}$ provides the characteristic lengthscale 
relevant to this regime, indicating the length on which
the magnetization undergoes a full $2\pi$~rotation in the absence of fields
and anisotropies. In Fig.~\ref{fig:phasesbyD} the helical length
can be estimated on the far right side as the distance between the centers
of neighboring skyrmions (or equivalently, from the distance between
midpoints of neighboring stripes of the same color). 

The vertical line for $D/D_{c}=1$ in Fig.~\ref{fig:phasesbyD}
separates the uniform from the helical phase. The film is placed
under an external field which penalizes the growth of the blue
magnetization regions and softens the transition from the uniform
state to the fully chiral phase. When $D/D_{c}<0.8$, the system is uniformly
magnetized. The region between the vertical array of stars and $D=1$
contains isolated skyrmions which survive because of topological protection,
with the skyrmions growing in size as $D$ increases. The configurations
found in the region $1<D/D_{c}<1.7$ are usually described as
skyrmion lattices, while for $D>1.7$ skyrmions coexist with a striped phase. 

It is important to note that the skyrmions shown in Fig.~\ref{fig:phasesbyD} are
stabilized by the external field; when this is decreased
below a certain value (the so-called strip-out field~\citep{leonovPropertiesIsolatedChiral2016}),
the system relaxes into the chiral phase. 

\begin{figure*}
\includegraphics[width=6in]{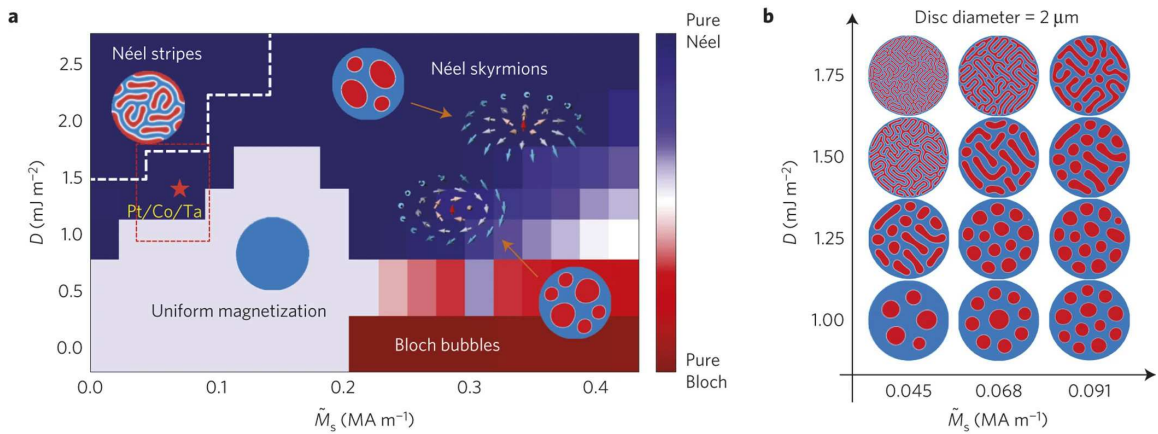}\caption{\label{fig:skyrmionstatesinnanodots} a) Phase diagram of ferromagnetic
nanodisks with various values of $D$ and $M$.  The boundary conditions
enforce an integral value of the topological number $\Omega$
for all states. b) Typical configurations for~$D\ge1$.  From~\citep{wooObservationRoomtemperatureMagnetic2016}.}
\end{figure*}

Skyrmions are further stabilized by geometric confinement~\citep{rohartSkyrmionConfinementUltrathin2013}, a consequence
of the boundary conditions at the flim's edge. These boundary
conditions depend mostly on the relative strength of the exchange interaction, DMI,
and surface anisotropies. This has motivated research
investigating the potential use of skyrmions as information carriers in
ferromagnetic nanodisks~\citep{sampaioNucleationStabilityCurrentinduced2013,winklerSkyrmionStatesDisk2021,guslienkoNeelSkyrmionStability2018,everschor-sittePerspectiveMagneticSkyrmions2018},
nanotracks~\citep{cortes-ortunoThermalStabilityTopological2017} and
other confined geometries~\citep{cortes-ortunoNanoscaleMagneticSkyrmions2019,novakMicromagneticStudySkyrmion2018}.
This research also includes theoretical studies of thermal stability~\citep{buttnerTheoryIsolatedMagnetic2018,hagemeisterStabilitySingleSkyrmionic2015,zelentBiStabilityMagneticSkyrmions2017}
as well as experimental observations~\citep{zelentSkyrmionFormationNanodisks2021}.
Just as with a decreasing field, reduction of the density of
skyrmions in a confined magnet causes them to decay into
meandering stripes, as shown in~Fig.~\ref{fig:skyrmionstatesinnanodots}.

\begin{figure}
\includegraphics[width=3in]{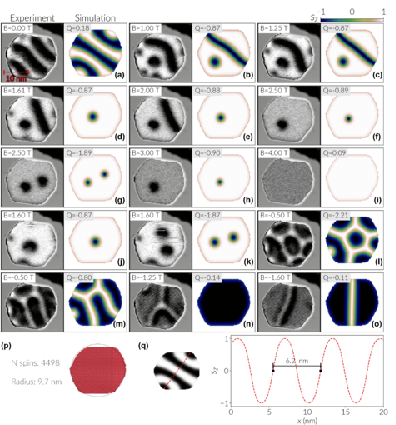}\caption{\label{fig:hexagonalislands} Equilibrium configurations at different
fields indicating their topological number (Q) in hexagonal nanoislands.
From~\citep{cortes-ortunoNanoscaleMagneticSkyrmions2019}.}
\end{figure}

Depending on boundary conditions, the topological number in confined
geometries need not be an integer. For example, Fig.~\ref{fig:hexagonalislands}
shows different states which appear in hexagonal nanoislands \citep{cortes-ortunoNanoscaleMagneticSkyrmions2019}.
By favoring the formation of domain walls,
energy landscapes of systems with large values of~$D$ exhibit
many energy minima with rich textures composed of skyrmion bubbles
and networks of domain walls. From this it becomes clear that knowing the presence or absence
of a skyrmion alone is not enough to characterize the magnetization texture.
Even for relatively small circular disks, the skyrmion becomes highly
asymmetric if $D$ becomes significantly large~\citep{kimBreathingModesConfined2014}.

In what follows, we will describe a different set of topological objects
which we believe are suitable candidates to characterize micromagnetic
configurations and which describe thermally activated transitions between
energy minima. As $D$ grows from zero and the phase changes from
the uniform configuration with isolated skyrmions into more complex
textures, the descriptors of the transition states must be modified
accordingly.  For an isolated skyrmion, such as those shown in Fig.~12, 
the topological number is an integer, and for $D>D_{c}$ it is stabilized by a magnetic field oriented
opposite to the direction of the magnetization at its core. If this
field is reduced below the strip-out field, the skyrmion becomes asymmetric~\citep{kimBreathingModesConfined2014} due to a competition between
the Zeeman energy (which penalizes the expansion of the core) and DMI
(which favors domain wall growth). 

As a result of this competition the skyrmion
grows in only one direction, forming a tubular structure with two semicircular
end caps, as shown in Fig.~16.
\begin{figure}
\includegraphics[width=3in]{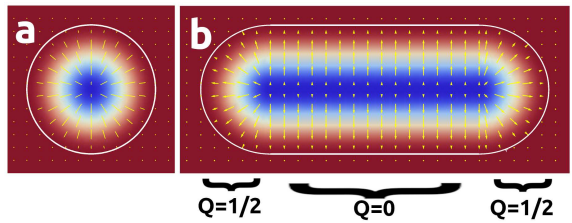}
\label{fig:skyrmionandbimerons}
\caption{(a) Skyrmion and (b) Bimeron}
\end{figure}
Each of the end caps is
called a meron~\citep{ezawaCompactMeronsSkyrmions2011}, which has half-integral winding number. A calculation
of the topological charge based on~(\ref{eq:winding}) shows that
the skyrmion number density is concentrated at the end caps, while
the central tube has zero topological density. The transition from
the skyrmion lattice to the helical phase can be interpreted as the
result of skyrmions stretching into a bi-meron
tube, with the end caps moving to infinity.   As we will see, merons are the key objects of thermal activation for large $D$ 
(beyond $D > 1.0$ in Fig.~13)


The last set of topological objects to be introduced is relevant for
large $D$, the helimagnetic phase. Fig.~\ref{fig:skyrmionstatesinnanodots}b
shows that a typical configuration in this regime contains intricate
labyrinthic paths with almost constant widths. In this regime, magnetic
singularities~\citep{klemanMagneticSingularitiesHelimagnetic1970,schoenherrTopologicalDomainWalls2018,fincoImagingTopologicalDefects2022}
prevent the system from relaxing to the cycloidal state with a constant
direction of the helical axis. These defects have many similarities
with disclinations in lamellar phases, and can be classified accordingly;
\begin{figure}
\includegraphics[width=3.5in]{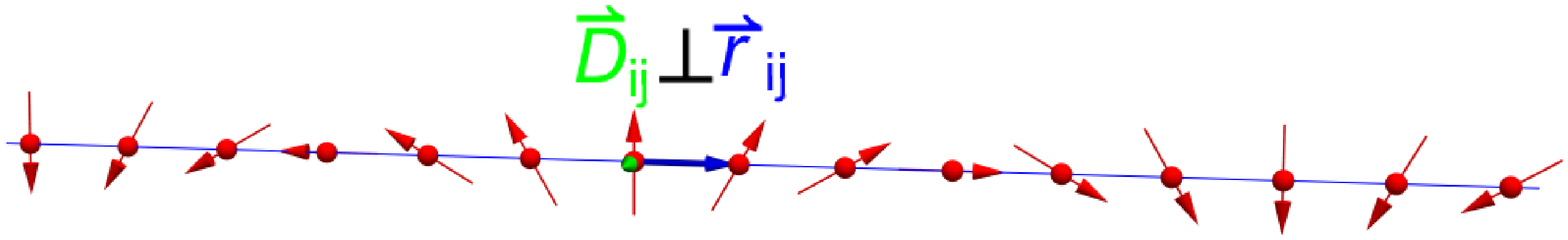}
\includegraphics[width=0.7in]{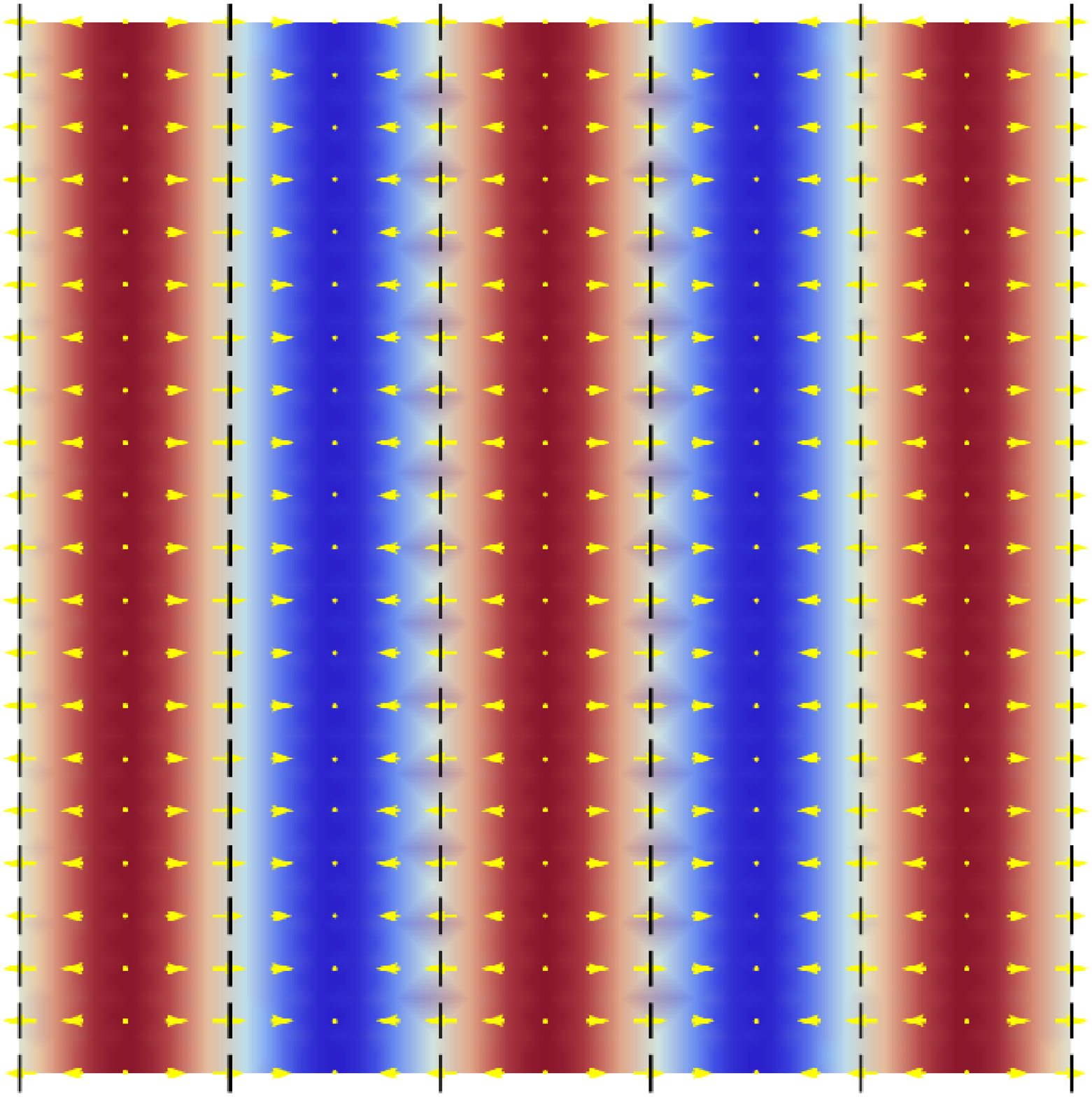}\includegraphics[width=0.7in]{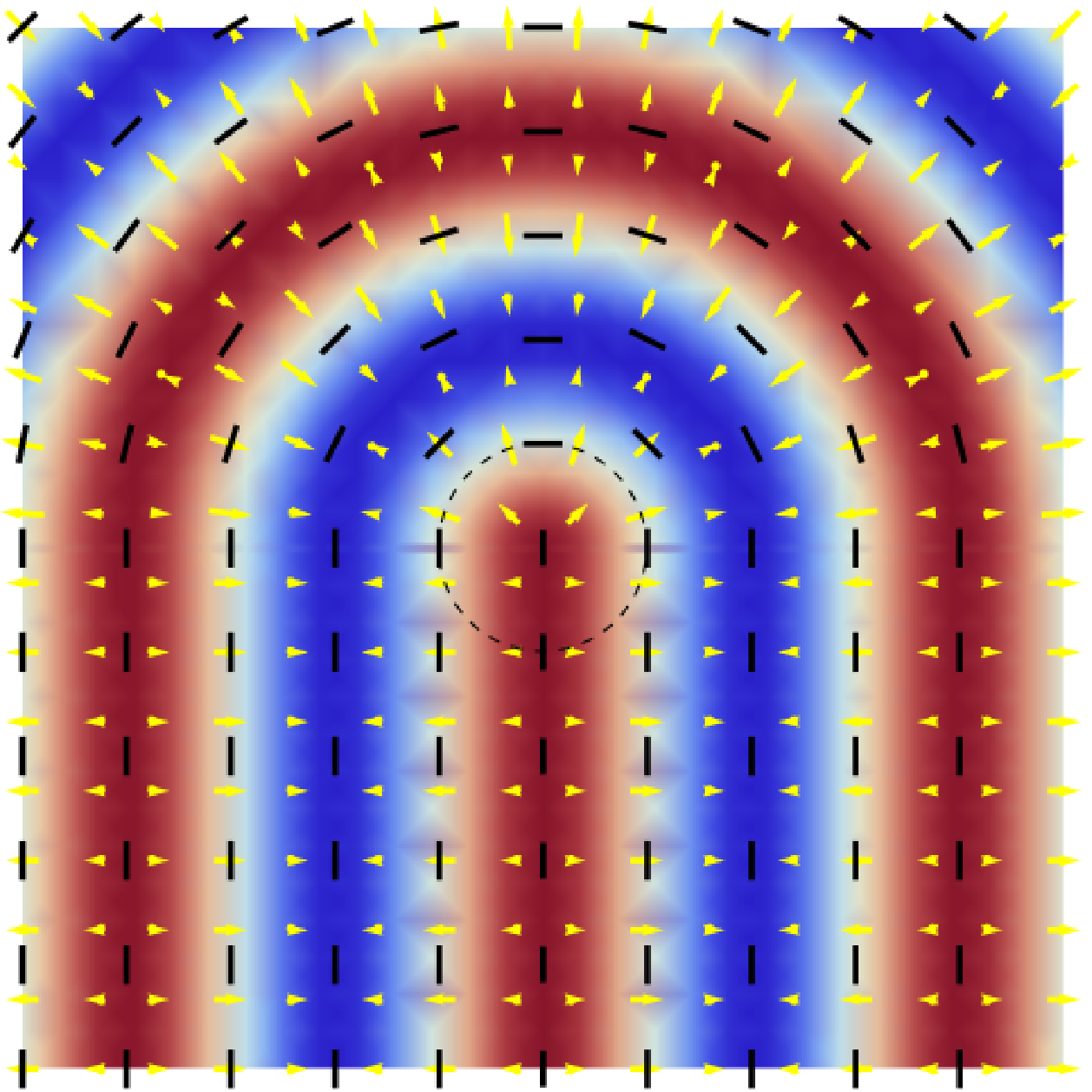}\includegraphics[width=0.7in]{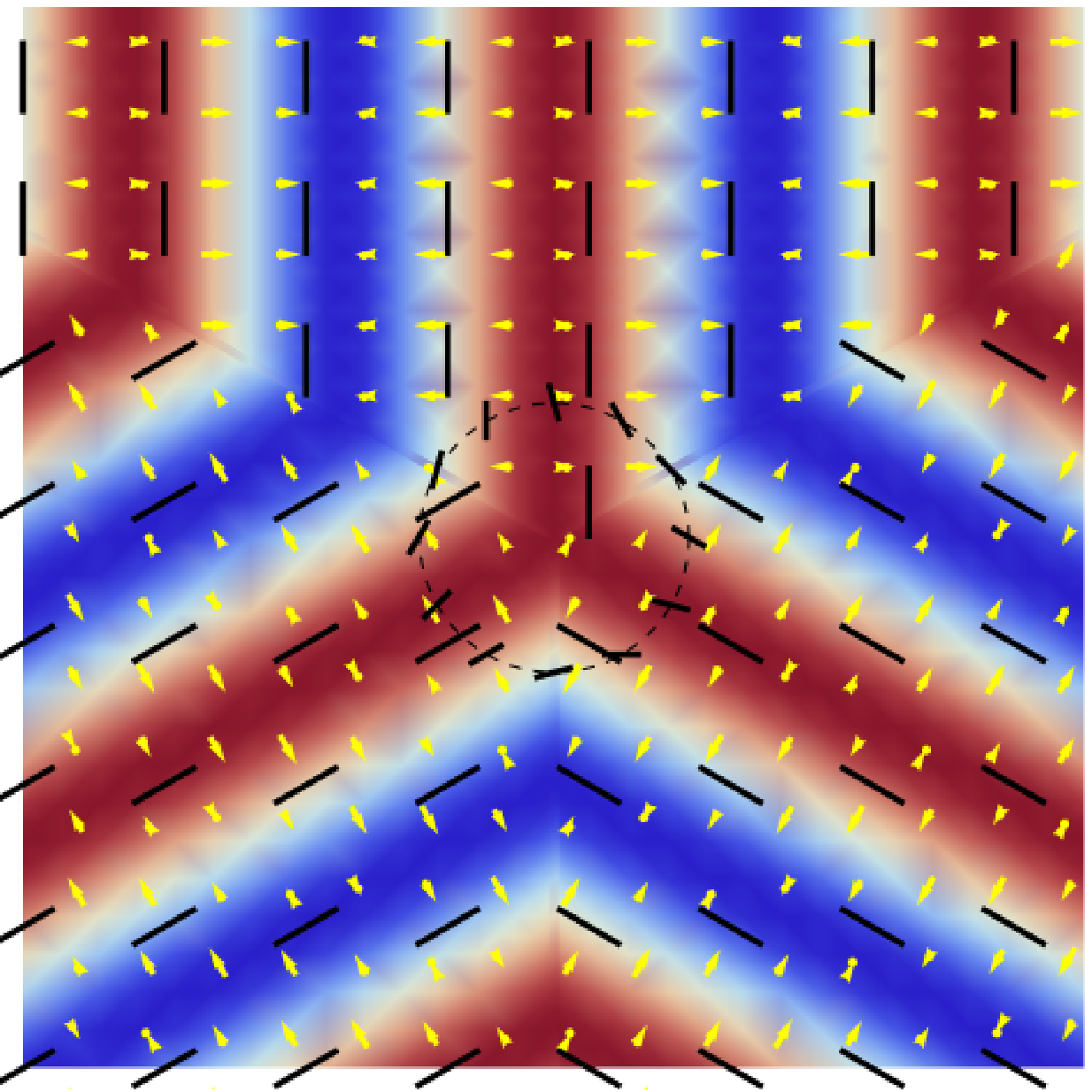}\includegraphics[width=0.7in]{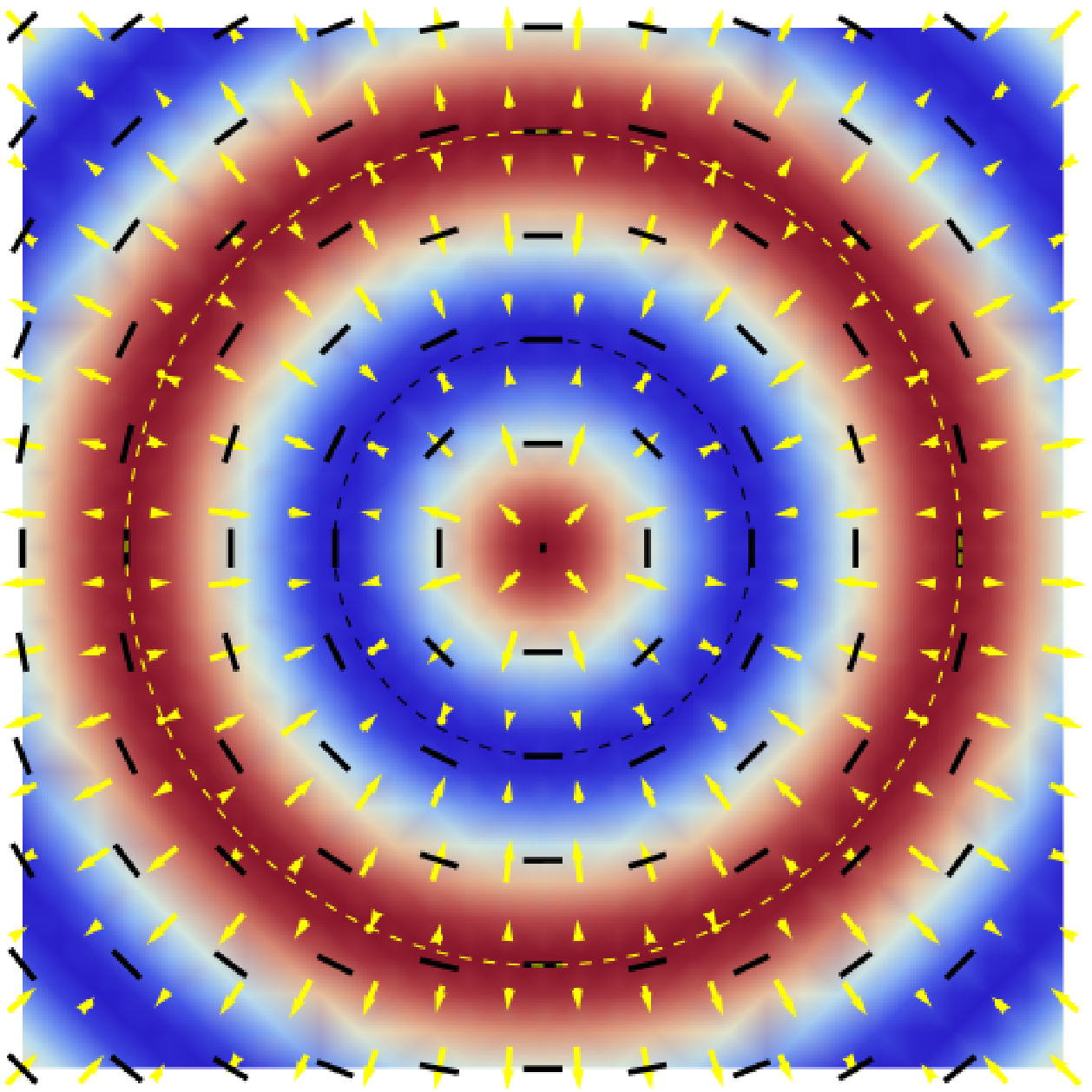}\includegraphics[width=0.7in]{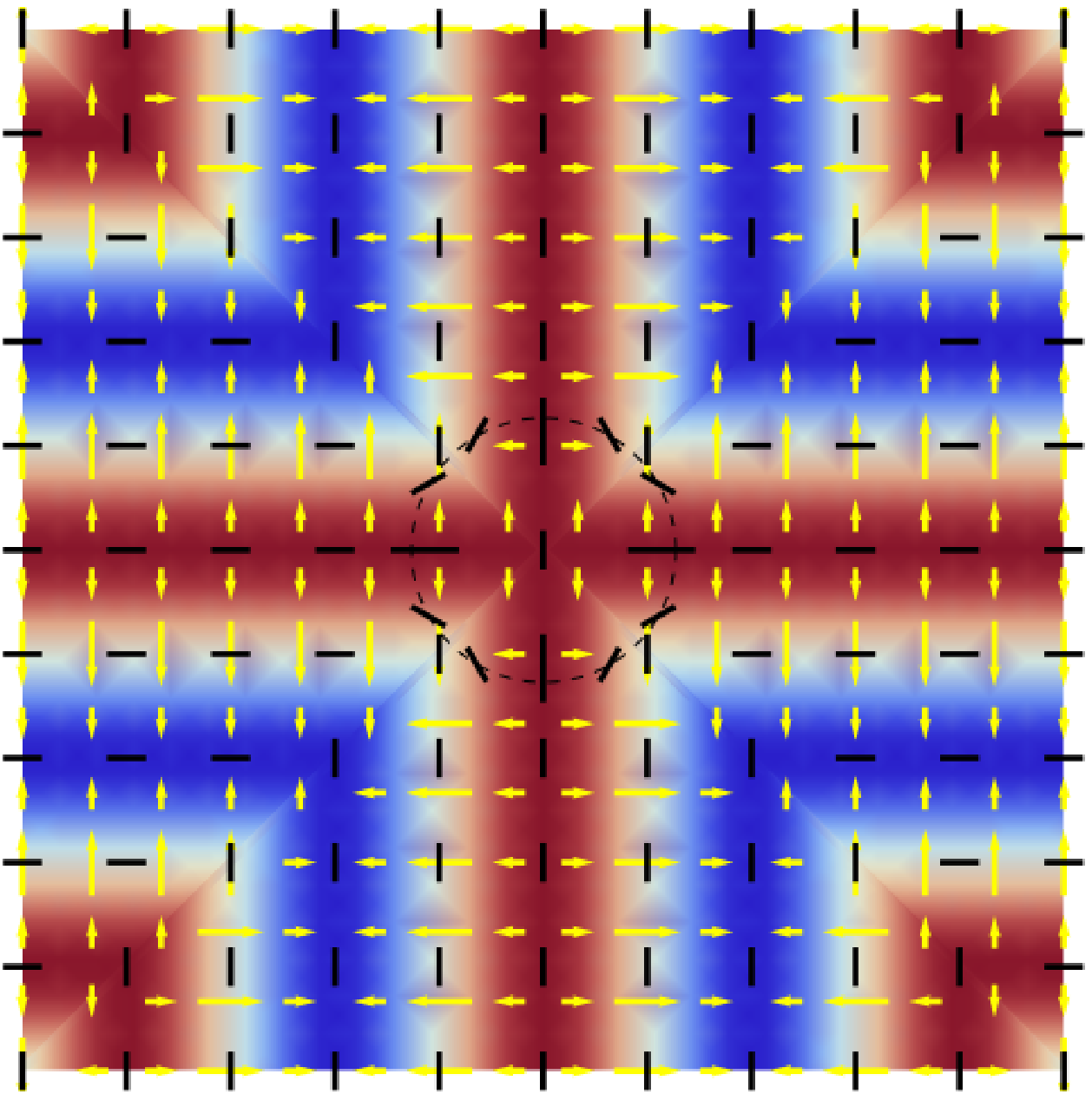}\caption{\label{fig:disclinations} (Top) Cycloid configurations of the magnetization
as a consequence of interfacial DMI that favors the formation of Neel
domain walls. The helical axis points out of the page and is perpendicular
to the vector that connects two neighboring atoms. (Bottom) Schematic
representations of typical structures. From left to right: (i) The
helical state; (ii) $+\pi$ disclination with the dashed circle encircling
a meron; (iii) $-\pi$ disclination; (iv) $+2\pi$ disclination with
a dashed black circle around a skyrmion and a yellow circle around
a ``skyrmionium particle''; (v) $-2\pi$ disclination. The yellow arrows indicate
the magnetization after projection to the $xy$~plane. The black segment
indicates the direction of the helical axis in different regions of
space. }
\end{figure}
they are shown in Fig.~\ref{fig:disclinations}. To explain the nomenclature,
we start by identifying the axis of rotation of the magnetization
in different regions of space. The value of the disclination is given
by how much the \textit{helical axis} winds for a path encircling
each object. The disclinations are then embedded in a helical
background as shown in Fig.~\ref{fig:embeddeddisclinations},
\begin{figure}
\includegraphics[width=3.5in]{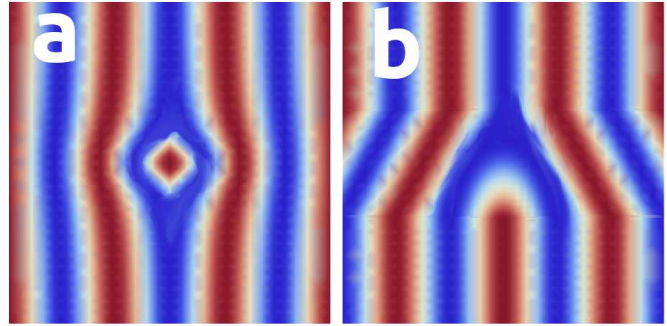}\caption{\label{fig:embeddeddisclinations} The isolated disclinations described
in Fig.~\ref{fig:disclinations} can be identified as the defects
we have previously introduced for moderate~$D$ . Here a skyrmion (a) and
a meron (b) are shown inside a helical background in two dimensions.}
\end{figure}
which incorporates previously introduced magnetic textures. With this it is now possible
to recognize the intricate structures shown in Fig.~\ref{fig:skyrmionstatesinnanodots}
as networks of disclinations connected by helical tubes.

\begin{figure}
\includegraphics[angle=-90,width=3.5in]{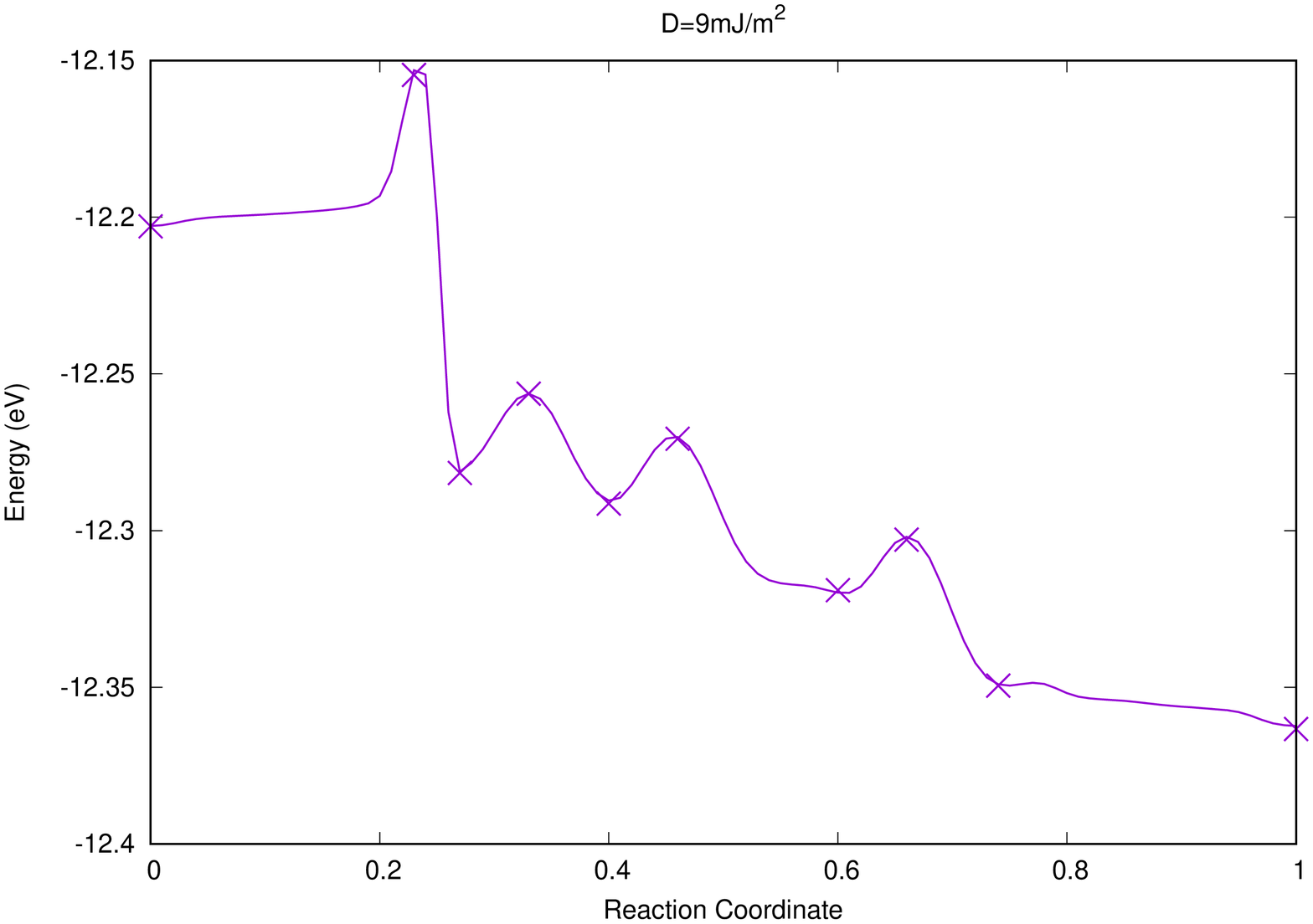}

\includegraphics[width=0.35in]{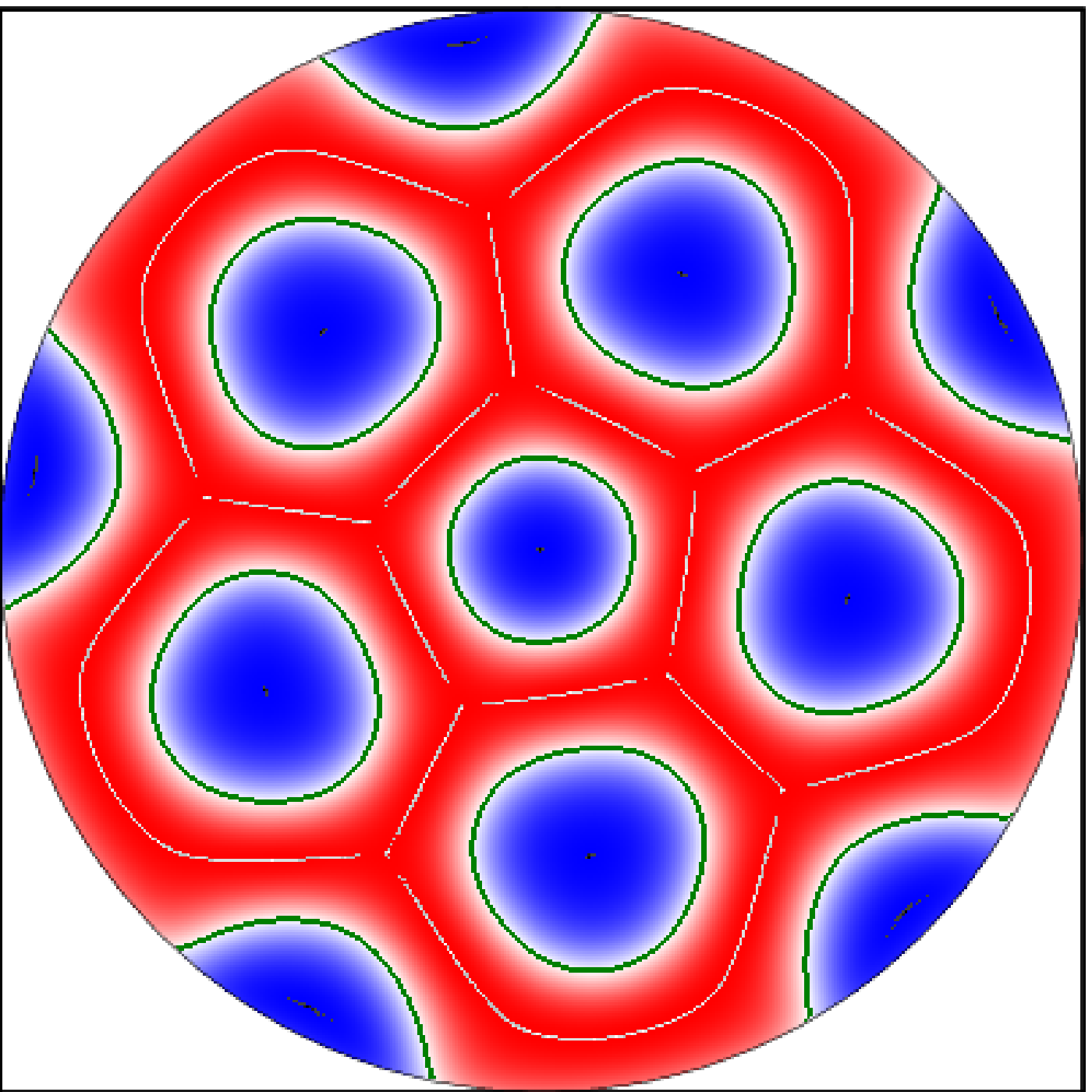}\includegraphics[width=0.35in]{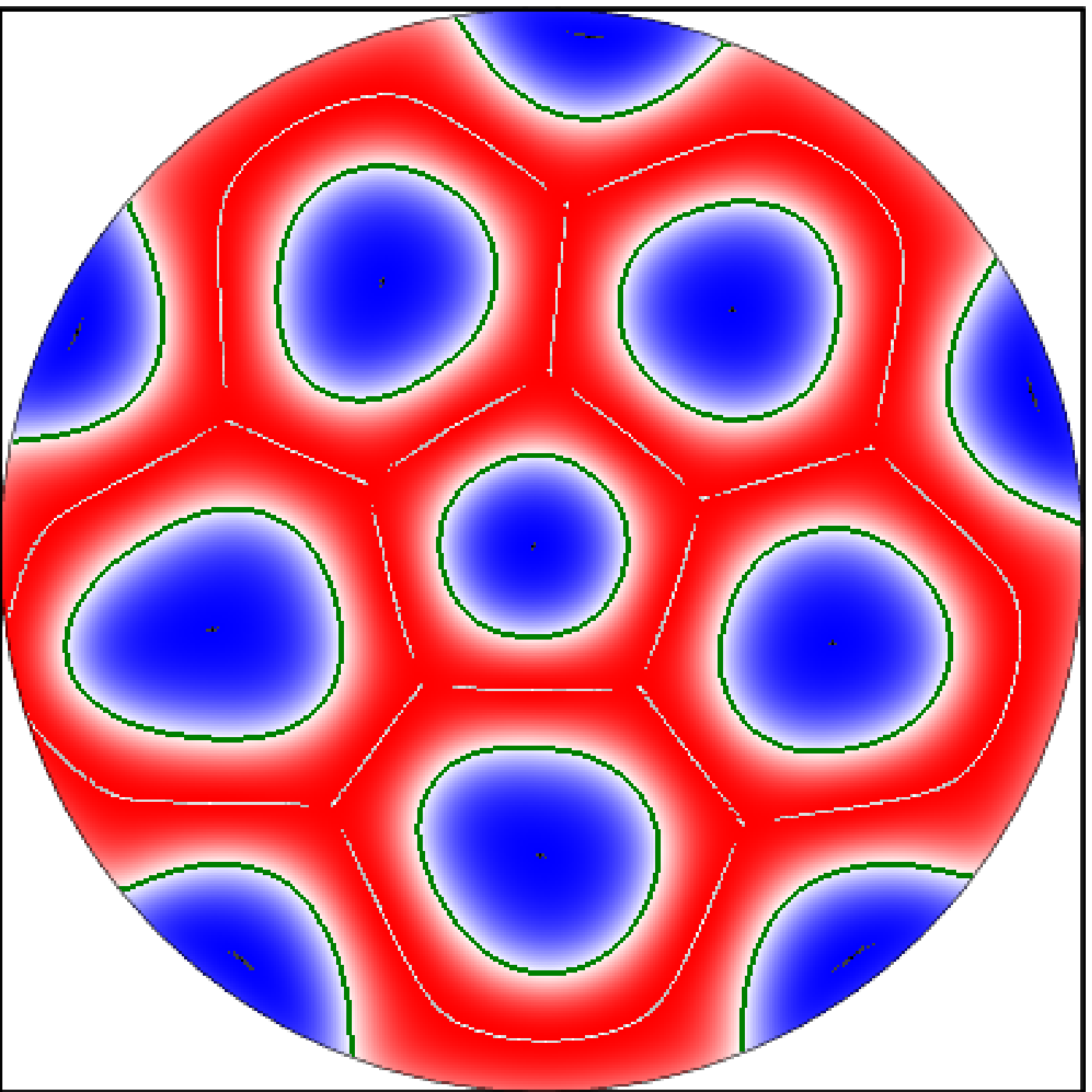}\includegraphics[width=0.35in]{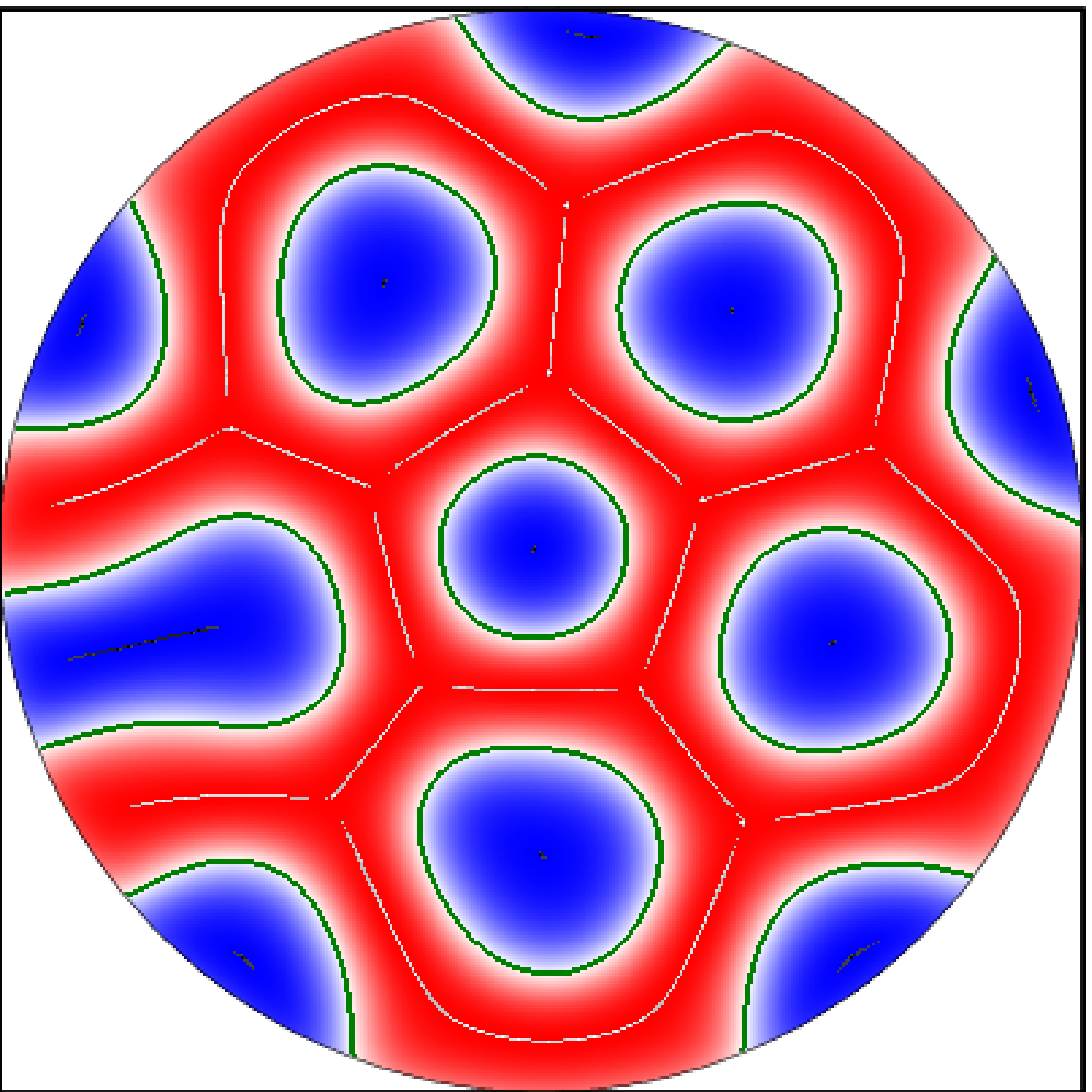}\includegraphics[width=0.35in]{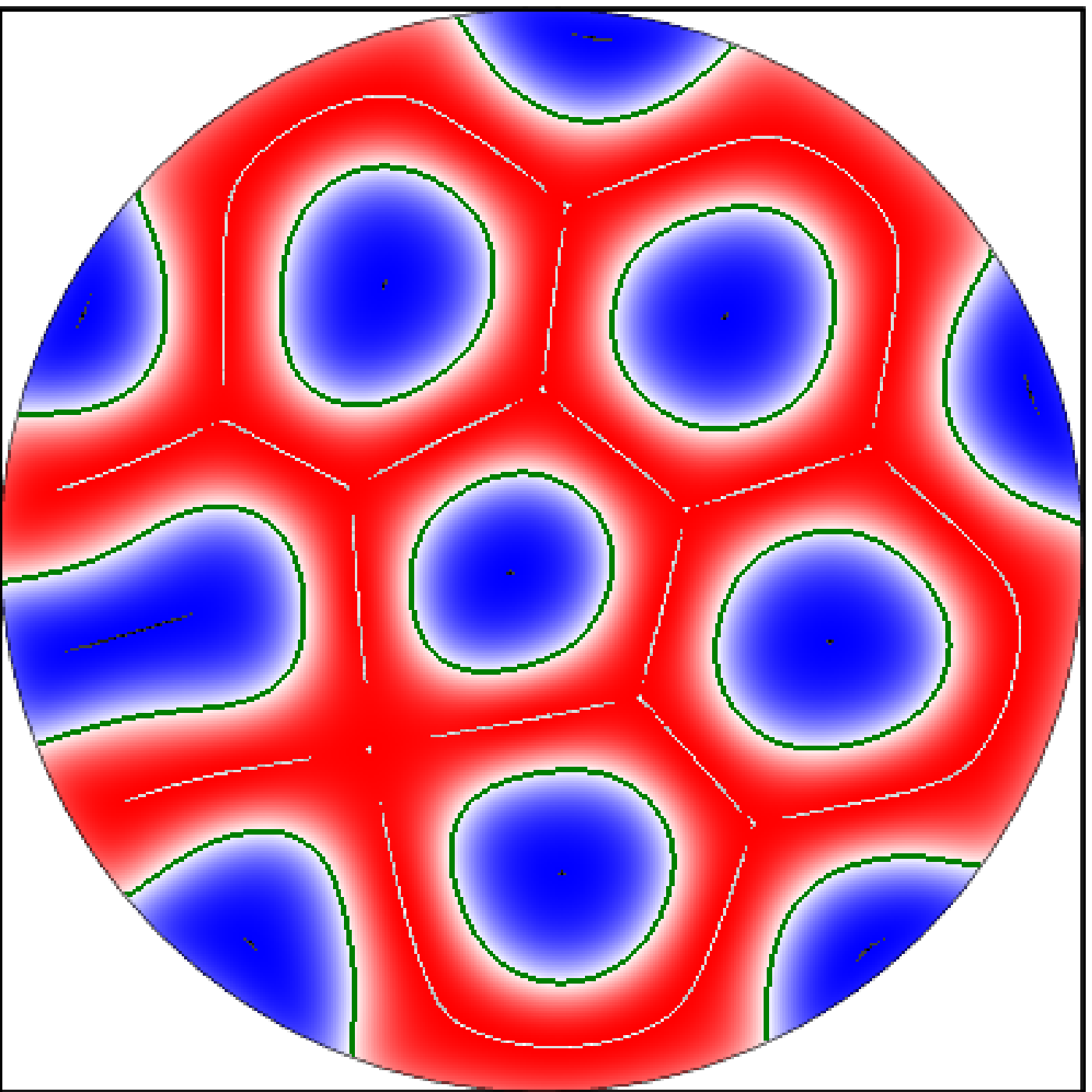}\includegraphics[width=0.35in]{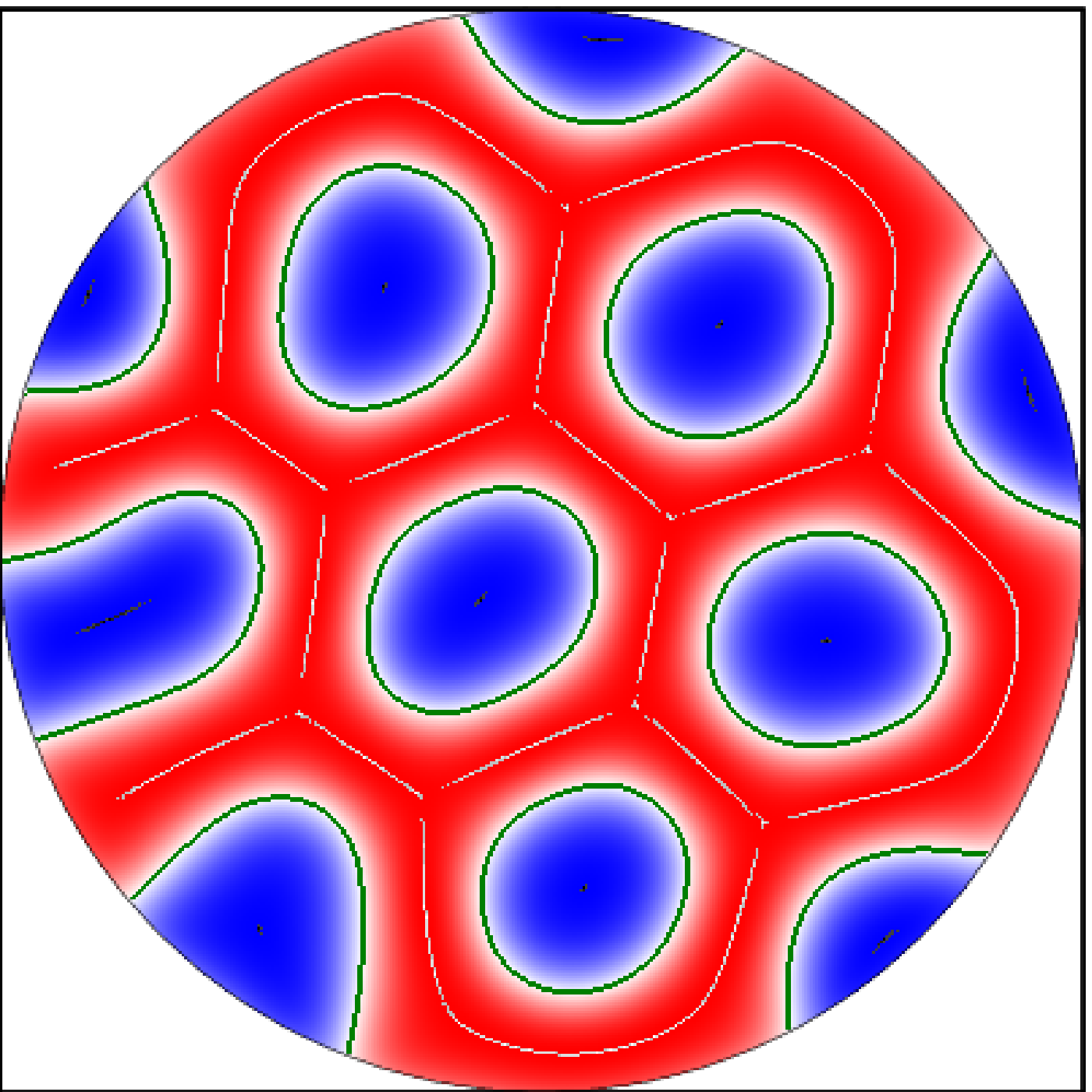}\includegraphics[width=0.35in]{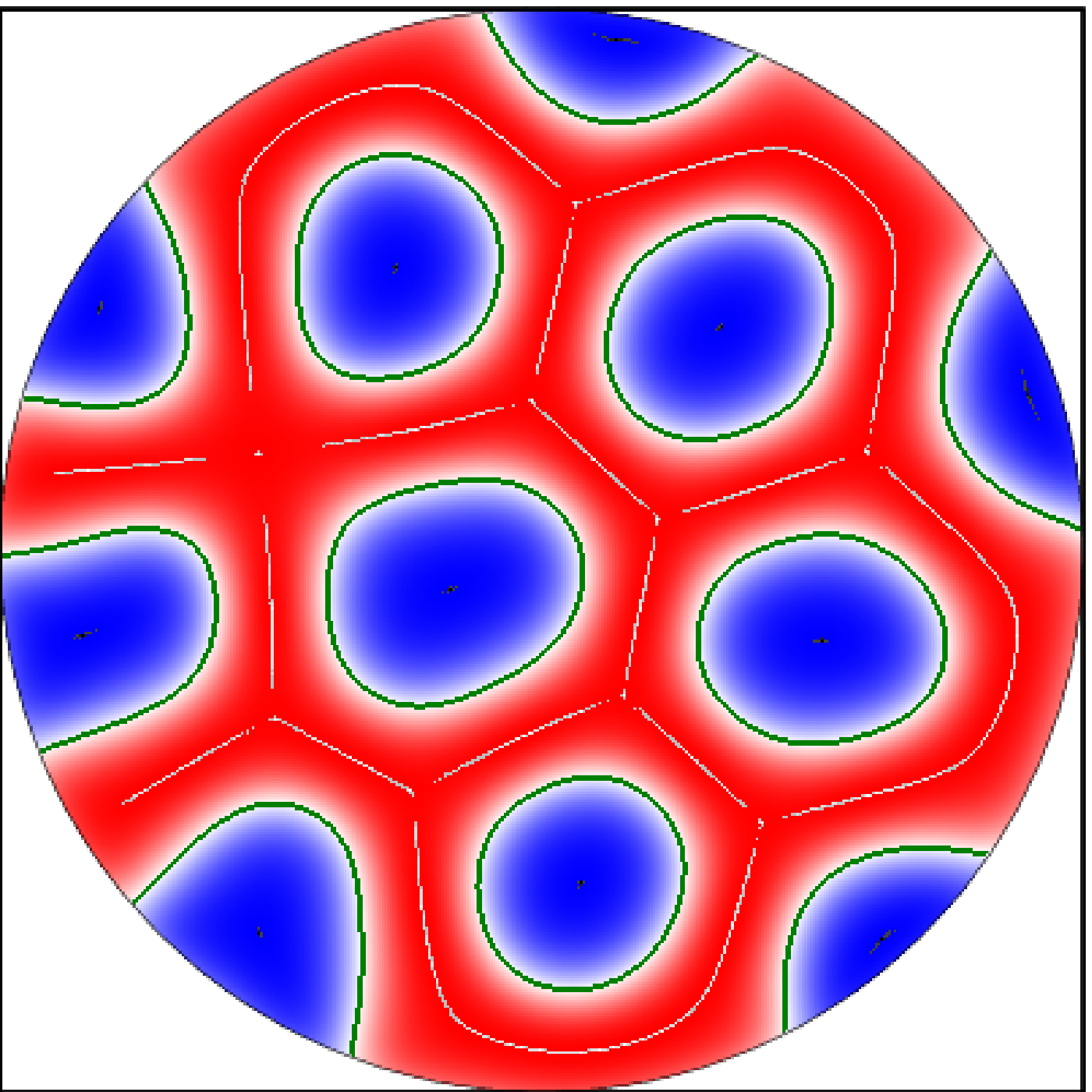}\includegraphics[width=0.35in]{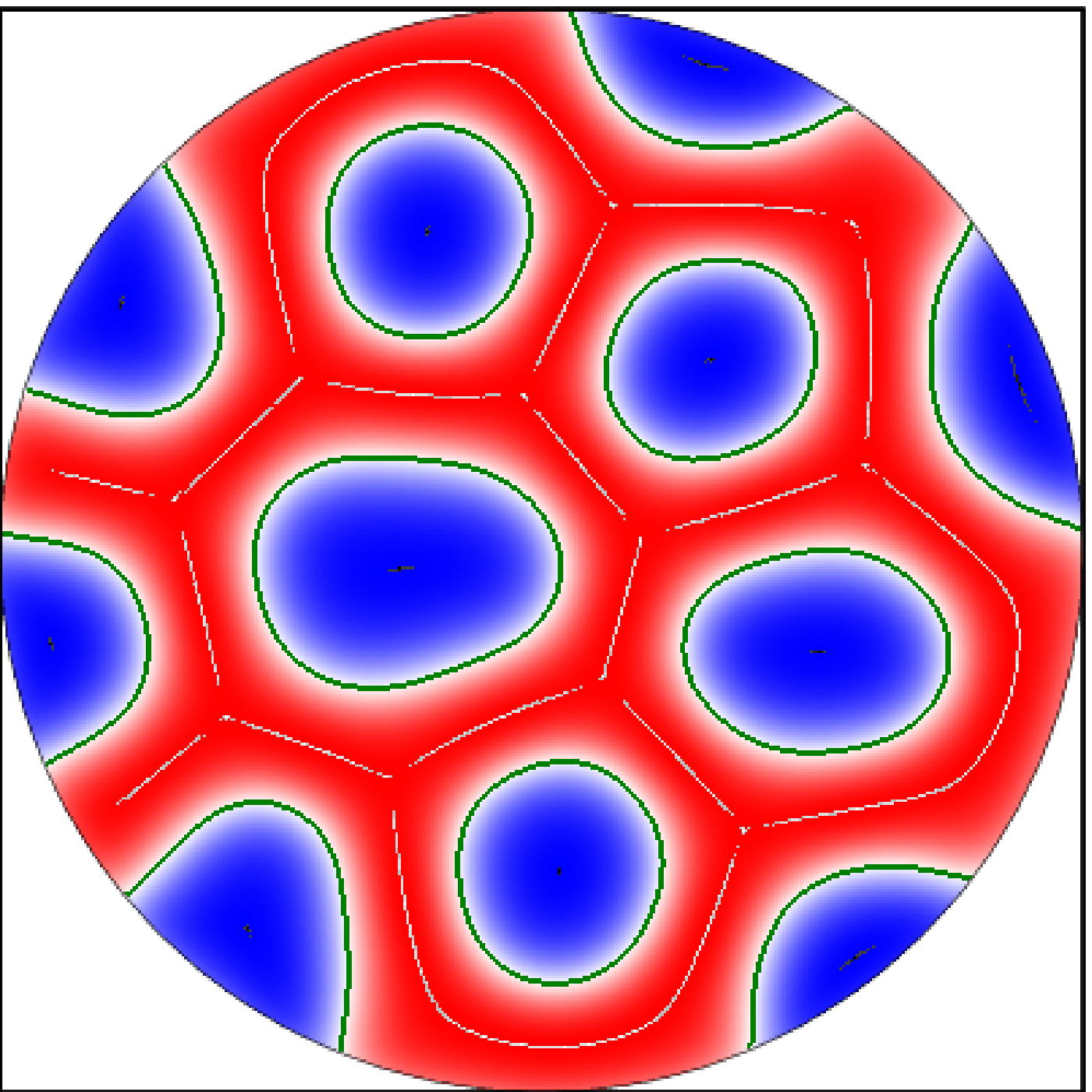}\includegraphics[width=0.35in]{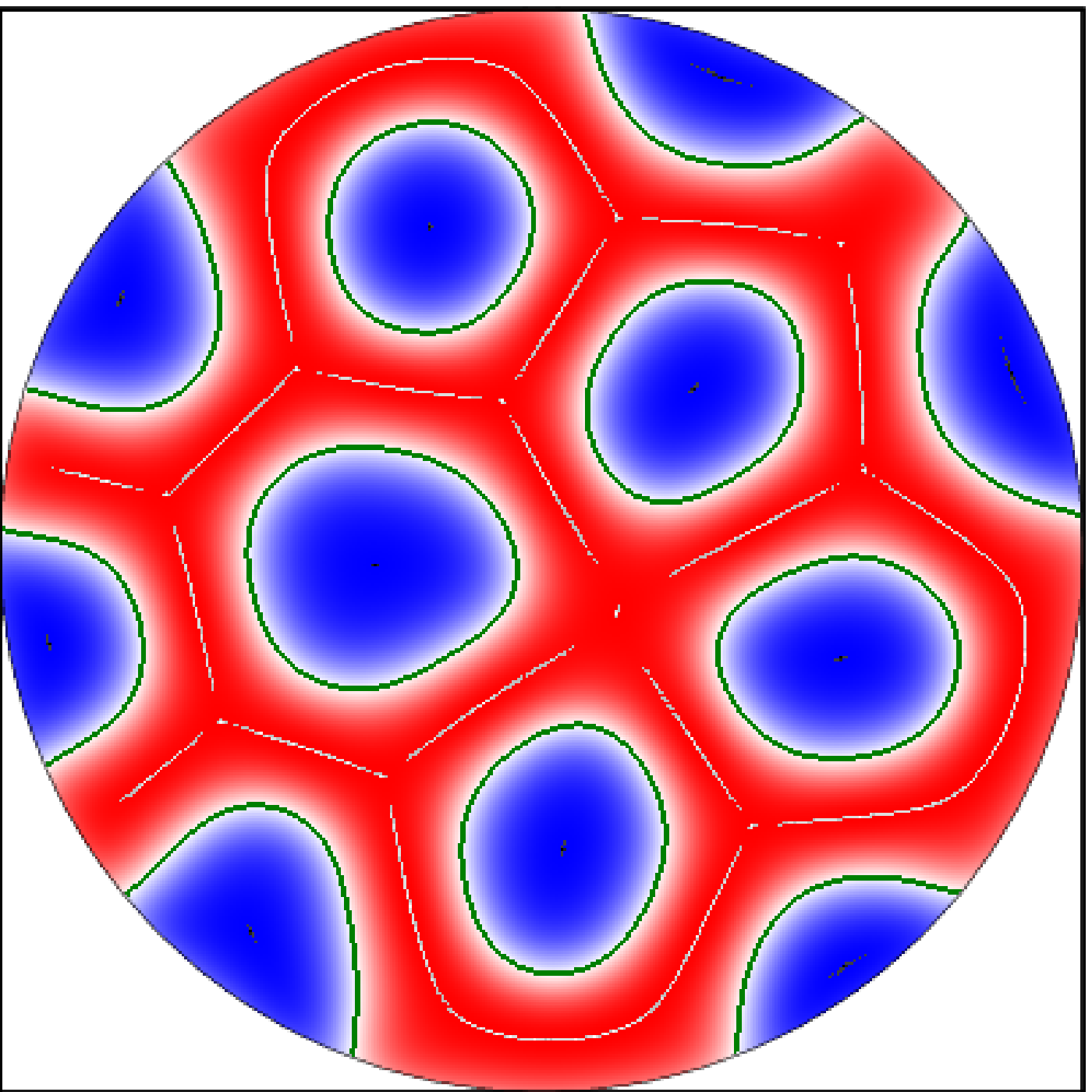}\includegraphics[width=0.35in]{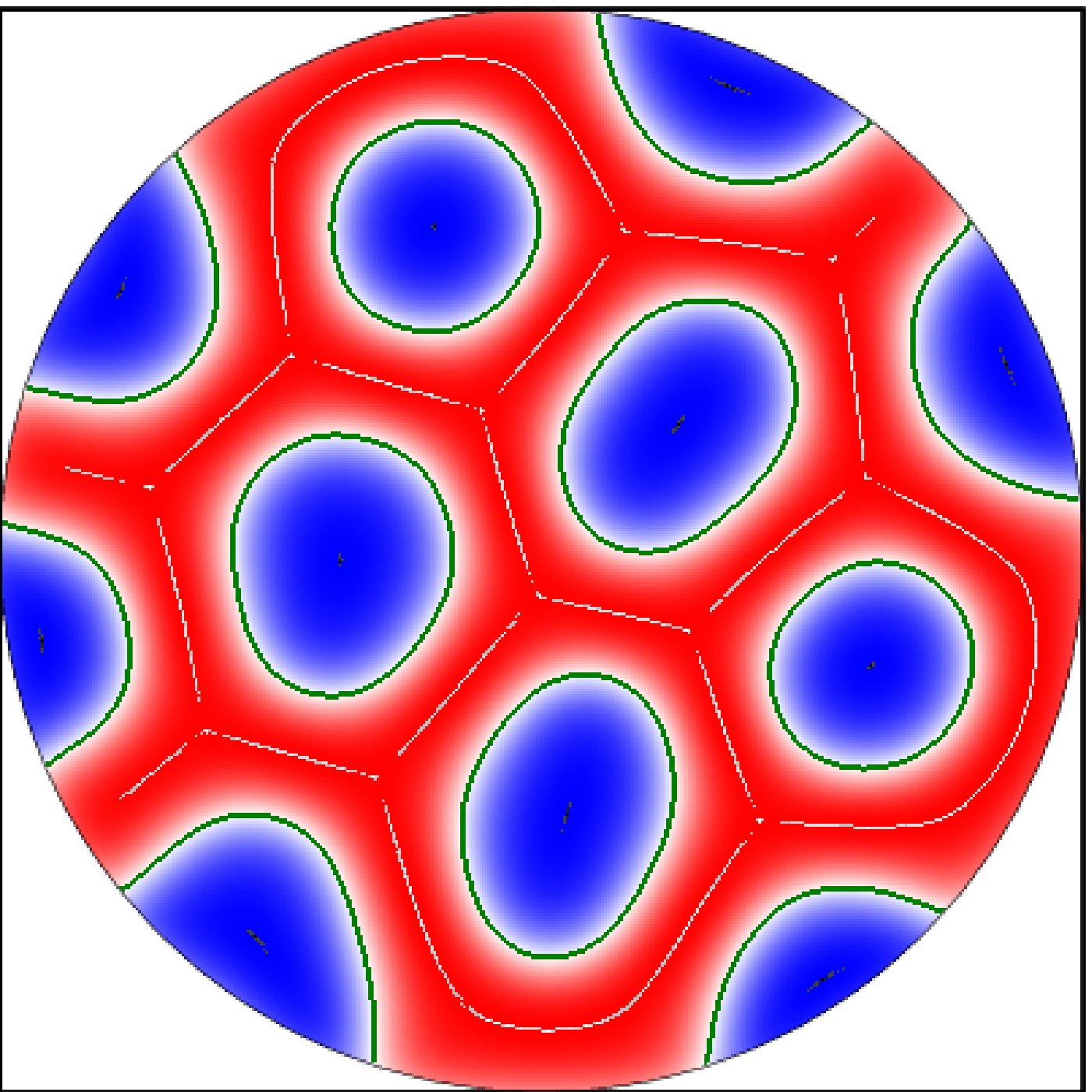}\includegraphics[width=0.35in]{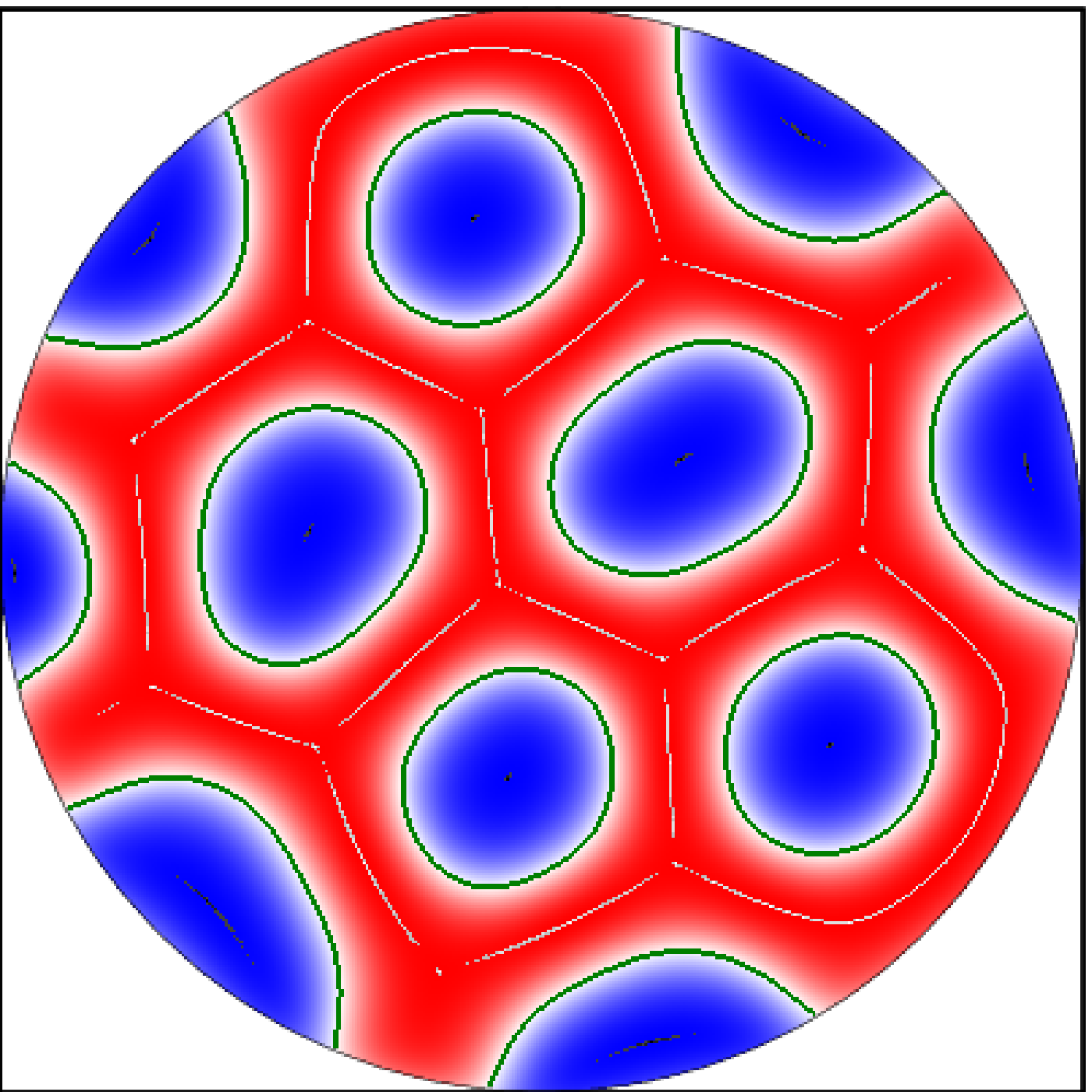}\caption{\label{fig:largeDtransitions} Results of string method calculations
for large~$D$.}
\end{figure}

We now present results of string method calculations of the energy
barrier for thermally activated transitions in a nanodisk with large $D$ ($=9$~mJ$\cdot$m$^{-3}$).
We selected two configurations that appeared to be close in configuration
space. The guessed path consisted of a continuous transformation between the final
and initial images (details will be provided elsewhere). The string
was allowed to slide down the energy landscape and the final result
is shown in Fig.~\ref{fig:largeDtransitions}: the graph shows the
energy as a function of the reaction coordinate and the figures beneath
correspond to the magnetization configuration at the points marked
in the graph. A surprising finding was that, although we expected
the basins of attraction of the initial and final states to be adjacent,
they were separated by multiple saddles. The first and highest maximum
corresponds to the expulsion of a meron at the edge. The rest of
the process consists of a reacommodation of the skyrmions. Maxima coincide
with situations in which two $-\pi$ disclinations are pushed together
and form a $-2\pi$ disclination (annihilating a domain wall). This
temporary $-2\pi$ disclination rapidly splits back into a pair of
$-\pi$ disclinations and a new wall. In consequence two skyrmionic
domains that were previously adjacent are now farther apart,
while two skyrmion domains that were far apart now become neighbors. 

The key takeaway from this calculation is that, as was the case with vortex-mediated annihilation of $2\pi$ domain walls, the thermally activated
reversal between micromagnetic states is mediated by magnetic singularities.
The reaccomodation of skyrmions inside a cluster is mediated by the merging
and breaking of disclinations, while skyrmions are destroyed by the
motion of merons across an edge. In future work our goal is to obtain appropriately
reduced descriptions that capture the essential dynamics of these
magnetic textures, and use it to estimate transition rates between
the different configurations.

\section{Summary}
\label{sec:summary}

In this paper we presented a mathematical formalism for understanding activation over
a barrier in micromagnetic systems in the presence of weak thermal noise. These systems
present a number of challenges in analyzing rare events leading to changes in magnetization configurations,
due to the presence of long-range interactions, finite-size effects, non-gradient dynamics, and the presence of
complex static or dynamical configurations as initial, final, and especially transition states.

In addition to employing analytical tools such as Wentzell-Freidlin path integrals and chain-of states
methods such as the string method, we have also introduced a general dynamical formalism and discussed useful techniques such as transformation
to a rotating frame (in the case of dynamical droplet soliton configurations) and collective coordinate methods.
These methods were used in part or in whole in analyzing transitions in several kinds of systems, and have often revealed a rich
energy landscape in the space of configurations. In particular, transitions between topologically protected structures such as magnetic skyrmions and
other configurations are mediated by a rich variety of textures, including edge textures (merons) with half-integral topological charge and
networks of disclinations.

\acknowledgments This research was supported in part by U.S. National
Science Foundation Grant DMR 1610416 (DLS) and
the National Science Centre Poland under OPUS funding
Grant No. 2019/33/B/ST5/02013 (GDC).

\bibliography{doeringpaper}

\end{document}